\documentclass{article}

\usepackage{PRIMEarxiv}

\usepackage[utf8]{inputenc}
\usepackage[T1]{fontenc}

\usepackage{amsthm,amsmath,amsfonts,amssymb}

\usepackage{booktabs}
\usepackage{array}
\usepackage{multirow}
\usepackage{adjustbox}

\usepackage{graphicx}
\graphicspath{{media/}}
\usepackage{subcaption}

\usepackage{algorithm}
\usepackage{algpseudocode}

\usepackage{tikz}
\usetikzlibrary{shapes.geometric,arrows.meta,positioning}

\usepackage{xcolor}
\usepackage{microtype}
\usepackage{nicefrac}
\usepackage{url}
\usepackage{caption}
\usepackage{natbib}
\usepackage{fancyhdr}
\usepackage{hyperref}

\newtheorem{definition}{Definition}
\newtheorem{theorem}{Theorem}
\newtheorem{corollary}{Corollary}
\newtheorem{remark}{Remark}

\title{Optimal Differentially Private Randomized Response Designs to Collect Sensitive Binary Data
}

\author{
  Bittu Karmakar \\
  \\
  Department of Mathematics \\
  Indian Institute of Technology Guwahati, India \\
  \\
   \And
    Palash Ghosh \\
    \\
Department of Mathematics \\
  Indian Institute of Technology Guwahati, India\\
  and \\
  Centre for Quantitative Medicine\\
  Duke-NUS Medical School\\
  National University of Singapore, Singapore \\
  \\
  Correspondence to: palash.ghosh@iitg.ac.in\\
}

\begin{document}
\maketitle

\begin{abstract}
Randomized response is a long-standing method for estimating the prevalence of sensitive attributes while protecting respondent privacy. It is increasingly used to generate synthetic binary data from real personal records, enabling data storage and sharing while protecting individual privacy. While surveys emphasize accurate estimation, synthetic data generation prioritizes privacy. Statisticians typically set sample sizes to achieve a target statistical power. However, we show that high power can increase the risk of privacy violations. We consider established randomized response designs with respect to statistical power and differential privacy, which quantifies the leakage of privacy. Our results reveal that common design strategies can yield either insufficient power or excessive privacy loss. We provide optimal parameter choices for randomized response models that jointly satisfy desired power and differential privacy constraints. We motivate and evaluate our approaches using a dataset from a randomized response survey conducted via Amazon Mechanical Turk on tax return misreporting, providing a policy-relevant testbed. Simulation studies evaluate the existence of optimal design parameters, identify designs that minimize the required sample size, and quantify sample size inflation relative to direct questioning. We provide a user-friendly web application (Shiny App) available at \url{https://iitg.ac.in/pgapps/DP_RR/} for designing randomized response studies to facilitate adoption.
\end{abstract}


\section{INTRODUCTION}

How many high school students in our city have ever used illegal drugs? What percentage of adults have provided misleading or incorrect information on their income tax returns? What proportion of people have had unprotected sexual intercourse with multiple partners in the past year? These seemingly straightforward questions pose a persistent challenge for survey researchers, policymakers, and public health officials. When confronted with direct inquiries about sensitive or stigmatized behaviors, such as substance use, tax evasion, or sexual practices that increase the risk of sexually transmitted infections, respondents can conceal the truth due to concerns about stigma, legal consequences, or confidentiality breaches. Even assurances of anonymity often fail to elicit honest answers, leading to systematic bias and unreliable prevalence estimates \citep{chang2023privacy, dwork2014algorithmic}. This challenge extends beyond traditional social and health sciences. In the digital era, technology companies face a similar dilemma when analyzing user behavior while upholding privacy standards. For example, a web browser developer might wish to identify how many users have enabled a new security feature or visited potentially harmful sites, but collecting such information directly could expose individuals’ browsing habits. To address this, companies like Google and Apple have developed privacy-preserving mechanisms that allow them to aggregate user data for trend analysis, product improvement, and threat detection, all without ever knowing an individual user’s choices or history \citep{erlingsson2014rappor, team2017learning}. By incorporating statistical techniques that introduce random noise or locally randomize responses on users' devices, these organizations can extract valuable aggregate insights while providing rigorous mathematical guarantees of individual privacy.

Randomized response (RR) designs offer a principled solution to the challenge of collecting sensitive binary (Yes/No) data without compromising respondent privacy \citep{cruyff2008accounting}. By incorporating a probabilistic (random) element into the response process, RR mechanisms ensure that an individual’s answer cannot be directly traced to their true status by the researcher or any third party. Figure \ref{fig:five_designs} describes five popular RR designs, which will be discussed later in detail. Suppose a researcher wants to estimate the proportion of students who have ever cheated on an exam. During the interview, each student privately spins a spinner labeled ``Cheated'' and ``Never Cheated'' (see Figure \ref{fig:warner}). The researcher asks, ``Have you ever cheated on an exam?'' If the spinner lands on the label that matches the student’s actual experience, the student answers ``Yes''; otherwise, they answer ``No'' \citep{warner1965randomized}. The researcher can’t tell who really cheated, but by knowing the probability of heads/tails, they can estimate the true cheating rate from the group responses while protecting individual privacy. This built-in uncertainty protects participants from potential repercussions, thereby encouraging greater honesty. Rather than relying on the respondent’s trust in assurances of confidentiality, the method guarantees plausible deniability at the level of each response. The key innovation of RR is that, while any single answer may be ambiguous, the aggregate data retains sufficient structure for valid statistical inference about the population as a whole \citep{greenberg1969unrelated, simmons1967unrelated, mangat1990alternative}. The flexibility of RR designs ranges from the use of coins and dice to more elaborate randomization schemes, enabling researchers to tailor the approach to the sensitivity of the question and the survey context \citep{chaudhuri2020randomized, blair2015design, kuk1990asking}. RR designs transform the fundamental obstacle of privacy protection into an integral part of the statistical design, improving data quality and enabling robust estimation of sensitive behaviors at the population level \citep{boruch1971assuring, fox1986randomized}.

\begin{figure}[p]
\centering

\begin{subfigure}{0.5\textwidth}
\centering
\resizebox{\linewidth}{!}{
\begin{tikzpicture}[
node distance=4.5cm,
every node/.style={font=\Huge\bfseries},
block/.style={ellipse, draw, align=center, line width=1.5pt,
minimum height=3cm},
decision/.style={diamond, draw, aspect=2, align=center, line width=1.5pt, minimum height=3cm},
arrow/.style={->}
]

\node (start) [block] {Interviewee spins\\ the spinner privately};
\node (dec) [decision, below=of start] {Spinner Result};

\node (A) [block, below left=2.8cm and 2.2cm of dec] {Lands on ``A''};
\node (B) [block, below right=2.8cm and 2.2cm of dec] {Lands on ``B''};

\node (A1) [block, below left=4cm and 1.6cm of A] {Is in Group A,\\ Response: ``Yes''};
\node (A2) [block, below right=4cm and -4.5cm of A] {Is in Group B,\\ Response: ``No''};

\node (B1) [block, below left=4cm and -4.5cm of B] {Is in Group A,\\ Response: ``No''};
\node (B2) [block, below right=4cm and 1.6cm of B] {Is in Group B,\\ Response: ``Yes''};

\draw [arrow] (start) -- (dec);
\draw [arrow] (dec) -- (A);
\draw [arrow] (dec) -- (B);
\draw [arrow] (A) -- (A1);
\draw [arrow] (A) -- (A2);
\draw [arrow] (B) -- (B1);
\draw [arrow] (B) -- (B2);

\end{tikzpicture}
}
\caption{Warner's design}
\label{fig:warner}
\end{subfigure}
\hfill
\begin{subfigure}{0.48\textwidth}
\centering
\resizebox{\linewidth}{!}{
\begin{tikzpicture}[
node distance=1.5cm,
every node/.style={font=\Huge\bfseries},
block/.style={ellipse, draw, align=center, line width=1.5pt,
minimum height=2cm},
decision/.style={diamond, draw, aspect=2, align=center, line width=1.5pt, minimum height=2cm},
arrow/.style={->}
]

\node (start) [block, minimum width=10cm] {Consider two decks of\\ cards, containing red\\ and black cards};
\node (dec) [decision, below=2cm of start] {Each deck of cards is\\ shuffled thoroughly};

\node (A) [block, below=2cm of dec] {The respondent draws one card\\ from each shuffled deck};

\node (A1) [block, below left=3.5cm and 0.45cm of A] {Is in Group A,\\ Response: Color of the card\\ from the first deck};
\node (A2) [block, below right=3.5cm and 0.45cm of A] {Not in Group A,\\ Response: Color of the card\\ from the second deck};

\draw [arrow] (start) -- (dec);
\draw [arrow] (dec) -- (A);
\draw [arrow] (A) -- (A1);
\draw [arrow] (A) -- (A2);

\end{tikzpicture}
}
\caption{Kuk's design}
\label{fig:kuk}
\end{subfigure}

\vspace{0.3cm}

\begin{subfigure}{0.48\textwidth}
\centering
\resizebox{\linewidth}{!}{
\begin{tikzpicture}[
node distance=2cm,
every node/.style={font=\Huge\bfseries},
block/.style={ellipse, draw, align=center, line width=1.5pt,
minimum height=3cm},
decision/.style={diamond, draw, aspect=2, align=center, line width=1.5pt, minimum height=3cm},
arrow/.style={->}
]

\node (start) [block, minimum width=5cm] {Roll a 6-sided Die};

\node (dec) [decision, below=of start] {Die Result};

\node (left) [block, below left=2.5cm and 5.5cm of dec, minimum width=3.5cm] {Lands on ``1''};
\node (mid) [block, below=1.8cm of dec, minimum width=4cm] {Lands on ``2'' or\\ ``3'' or ``4'' or ``5''};
\node (right) [block, below right=2.5cm and 5.5cm of dec, minimum width=3.5cm] {Lands on ``6''};

\node (lresp) [block, below=of left, xshift=-1.7cm, minimum width=3.5cm] {Respond ``No''};
\node (mresp) [block, below=of mid, minimum width=3.5cm] {Respond ``Truthfully''};
\node (rresp) [block, below=of right, xshift=1.8cm, minimum width=3.5cm] {Respond ``Yes''};

\draw [arrow] (start) -- (dec);

\draw [arrow] (dec) -- (left);
\draw [arrow] (dec) -- (mid);
\draw [arrow] (dec) -- (right);

\draw [arrow] (left) -- (lresp);
\draw [arrow] (mid) -- (mresp);
\draw [arrow] (right) -- (rresp);

\end{tikzpicture}
}
\caption{Forced response design}
\label{fig:forced}
\end{subfigure}
\hfill
\begin{subfigure}{0.48\textwidth}
\centering
\resizebox{\linewidth}{!}{
\begin{tikzpicture}[
node distance=1.5cm, 
every node/.style={font=\Huge\bfseries}, 
block/.style={
ellipse,
draw,
align=center,
line width=1.5pt,
minimum height=2cm,
minimum width=3cm,
inner ysep=12pt
},
decision/.style={
diamond,
draw,
aspect=2,
align=center,
line width=1.5pt,
minimum height=2cm,
inner ysep=10pt
},
arrow/.style={->, line width=1.5pt}
]

\node (start) [block, minimum width=9cm]
{Interviewee spins the \\spinner privately};

\node (dec) [decision, below=of start]
{Spinner Result};

\node (A) [block, below left=3cm and 2cm of dec, minimum width=6.5cm, minimum height=2cm]
{Lands on ``A''};

\node (Y) [block, below right=3cm and 2cm of dec, minimum width=6.5cm, minimum height=2cm]
{Lands on ``Y''};

\node (Aresp) [block, below=4cm of A, xshift=-2.1cm, minimum width=7cm]
{Answer the sensitive question:\\ ``Yes'' or ``No''};

\node (Yresp) [block, below=4cm of Y, xshift=2.1cm, minimum width=7cm]
{Answer the unrelated question:\\ ``Yes'' or ``No''};

\draw [arrow] (start) -- (dec);
\draw [arrow] (dec) -- (A);
\draw [arrow] (dec) -- (Y);
\draw [arrow] (A) -- (Aresp);
\draw [arrow] (Y) -- (Yresp);

\end{tikzpicture}
}
\caption{Unrelated question design}
\label{fig:unrelated}
\end{subfigure}

\vspace{0.3cm}

\begin{subfigure}{0.6\textwidth}
\centering
\resizebox{\linewidth}{!}{
\begin{tikzpicture}[
node distance=1.5cm,
every node/.style={font=\Huge\bfseries},
block/.style={ellipse, draw, align=center, line width=1.5pt,
minimum height=2.2cm},
decision/.style={diamond, draw, aspect=2, align=center, line width=1.5pt, minimum height=3cm},
arrow/.style={->}
]

\node (start) [block, minimum width=5cm] {Flip a coin};
\node (dec1) [decision, below=of start] {Coin Flip Result};

\node (H1) [block, below left=2cm and 3cm of dec1, minimum width=4cm] {Lands on ``Heads''};
\node (T1) [block, below right=2cm and 3cm of dec1, minimum width=4cm] {Lands on ``Tails''};

\node (truth) [block, below=of H1, minimum width=4.5cm] {Respond ``Truthfully''};

\node (flip2) [block, below=of T1, minimum width=4.5cm] {Flip a second coin};
\node (dec2) [decision, below=of flip2] {Coin Flip Result};

\node (H2) [block, below left=2cm and 2.5cm of dec2, minimum width=4cm] {Lands on ``Heads''};
\node (T2) [block, below right=2cm and 2.5cm of dec2, minimum width=4cm] {Lands on ``Tails''};

\node (yes) [block, below=of H2, minimum width=4cm] {Respond ``Yes''};
\node (no)  [block, below=of T2, minimum width=4cm] {Respond ``No''};

\draw [arrow] (start) -- (dec1);

\draw [arrow] (dec1) -- (H1);
\draw [arrow] (dec1) -- (T1);

\draw [arrow] (H1) -- (truth);

\draw [arrow] (T1) -- (flip2);
\draw [arrow] (flip2) -- (dec2);

\draw [arrow] (dec2) -- (H2);
\draw [arrow] (dec2) -- (T2);

\draw [arrow] (H2) -- (yes);
\draw [arrow] (T2) -- (no);

\end{tikzpicture}
}
\caption{Two-Step design}
\label{fig:twostep}
\end{subfigure}

\caption{Schematic diagrams of five Randomized Response designs showing how randomized answers are generated.}
\label{fig:five_designs}
\end{figure}

Data has become a cornerstone asset in the modern era, intensifying the conflict between protecting individual privacy and maintaining the usefulness of data for analysis. While randomized response offers privacy protections tailored to survey settings, the broader landscape of data analysis, encompassing large-scale databases, administrative records, and digital traces, demands formal, generalizable standards for privacy. Differential privacy, introduced by \cite{dwork2006differential}, provides such a standard by furnishing a rigorous mathematical framework that quantifies the privacy risk associated with any data release. Under differential privacy, the presence or absence of any single individual (or record) has a provably limited effect on the results of an analysis, thus drastically reducing the risk of re-identification or the leakage of sensitive information. This framework has rapidly gained traction across computer science, statistics, and data science, and is now deployed in synthetic data generation, privacy-preserving machine learning, and the public release of aggregate statistics. Major organizations, including Google \citep{erlingsson2014rappor}, Apple \citep{team2017learning}, Microsoft \citep{ding2017collecting}, and the US Census Bureau \citep{dajani2017modernization}, have adopted differential privacy to enable valuable data analysis while safeguarding individual confidentiality. The widespread adoption and robust theoretical guarantees of differential privacy make it a cornerstone of modern data protection, offering practical solutions and a clear standard in an era of increasing data sensitivity. Therefore, integrating differential privacy into randomized response designs is essential, as it provides a rigorous and quantifiable privacy guarantee that strengthens the confidentiality of individual responses, even in the presence of sophisticated inference attacks or auxiliary information.

\begin{figure*}[htbp]
\centering

\begin{subfigure}{0.3\textwidth}
    \centering
    \includegraphics[width=\linewidth]{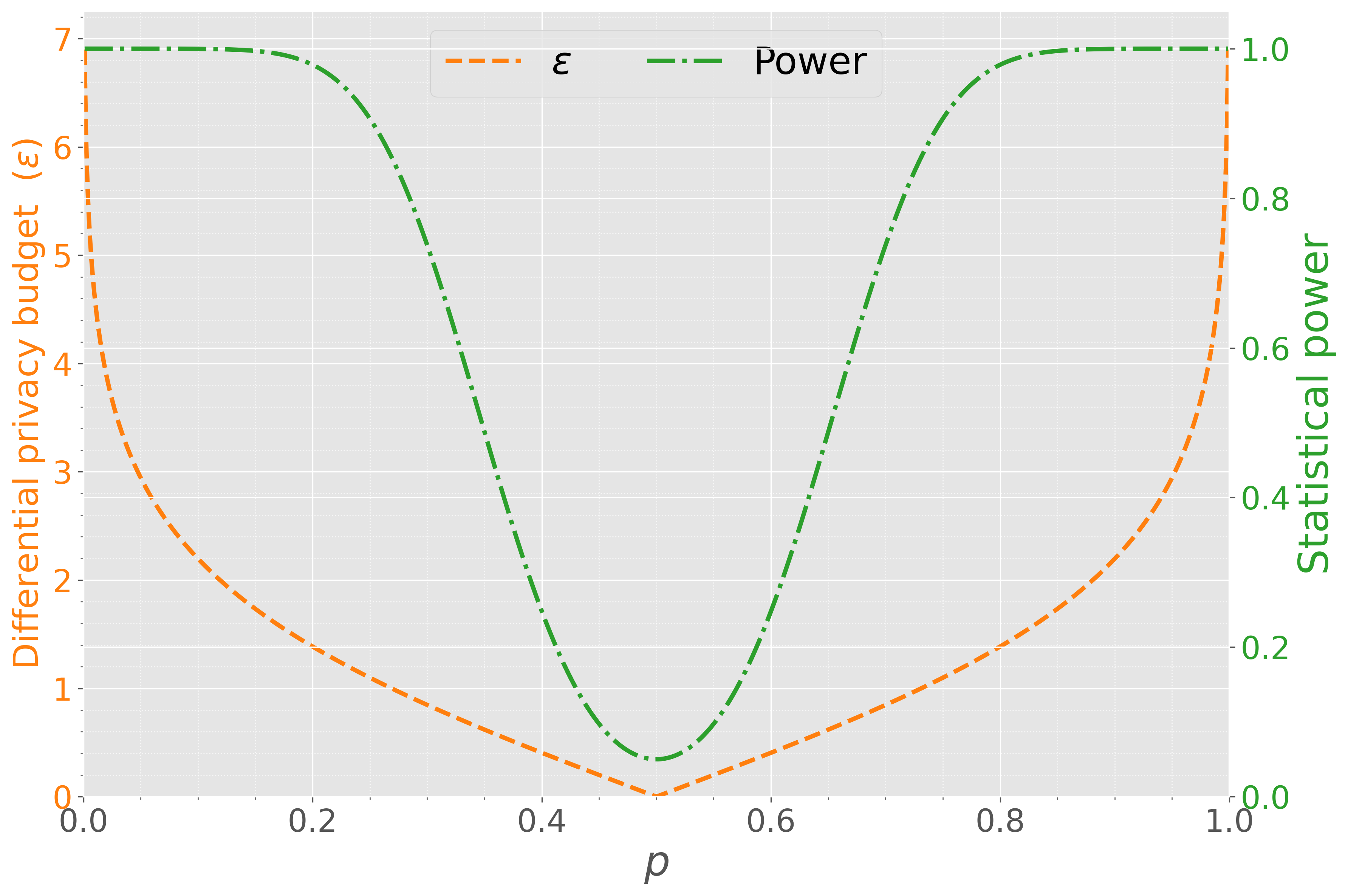}
    \caption{Warner}
    \label{fig:warner_com}
\end{subfigure}\hfill
\begin{subfigure}{0.3\textwidth}
    \centering
    \includegraphics[width=\linewidth]{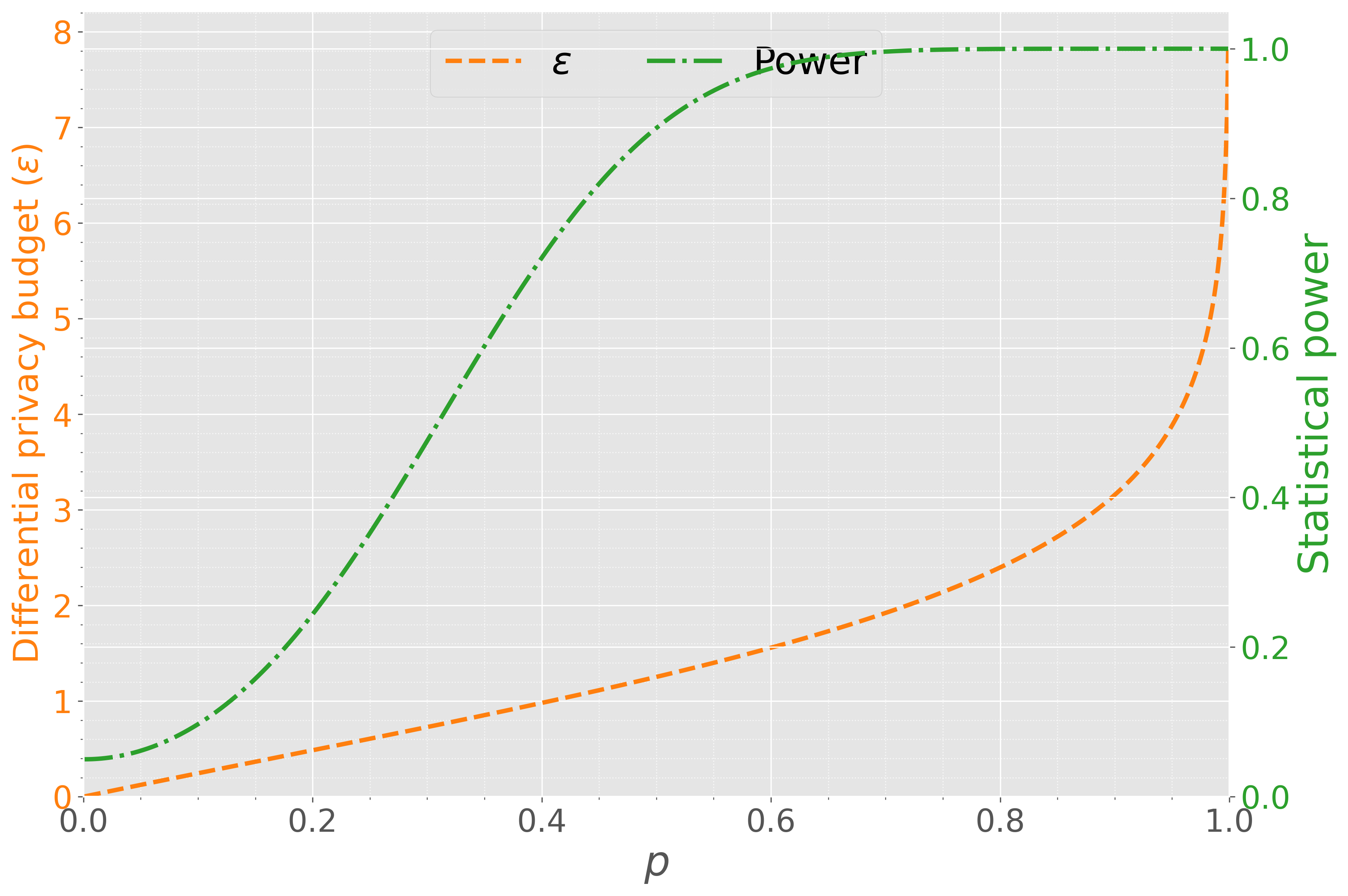}
    \caption{Unrelated}
    \label{fig:unrelated_com}
\end{subfigure}\hfill
\begin{subfigure}{0.3\textwidth}
    \centering
    \includegraphics[width=\linewidth]{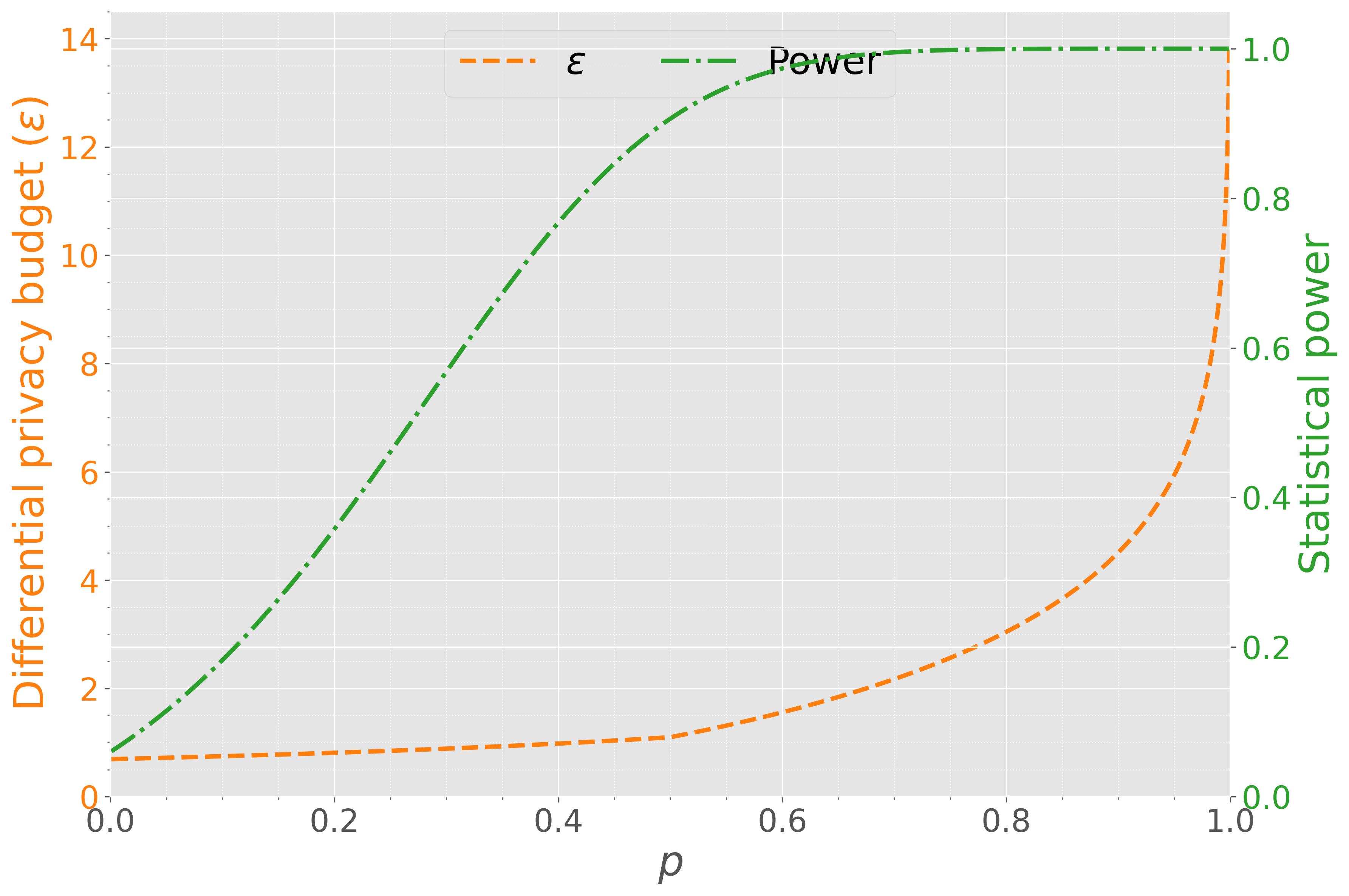}
    \caption{Two-Step}
    \label{fig:two_step_com}
\end{subfigure}

\caption{Comparison of statistical power and privacy budget for three randomized response designs.}
\label{fig:rr_comparison10}

\end{figure*}

In randomized response designs, the principal objective from a traditional statistical perspective is to maximize the accuracy of prevalence estimates (how many students have ever used illegal drugs?), typically by optimizing statistical power and minimizing estimation error \citep{chaudhuri2020randomized}. Historically, privacy in surveys using RR design has been addressed through the inherent masking provided by randomization, with less emphasis on formal mathematical guarantees and their quantification. However, the development of differential privacy in the computer science literature has shifted the focus toward rigorous, quantifiable privacy protections, ensuring that participation in a survey does not meaningfully increase an individual's risk of disclosure, even in the presence of auxiliary information (in reidentification). However, these two objectives are naturally in conflict. When we increase privacy in RR designs, for example, by setting a stricter differential privacy budget (denoted by $\epsilon$, a lower value is better), we have to add more noise to each response. This extra noise makes the collected data less accurate and reduces the statistical power of any analysis. Conversely, designs that prioritize statistical accuracy by allowing less noise inevitably weaken individual privacy guarantees (higher values of $\epsilon$). This privacy-utility trade-off is a fundamental challenge in the design of RR mechanisms under differential privacy, and is well recognized in the literature \citep{dwork2014algorithmic, kairouz2014extremal, wang2016using}. Using three well-known randomized response designs (Warner's Design, the Unrelated Question Design, and the Two-Step Design, which will be discussed in detail later), we illustrate this trade-off in Figure \ref{fig:rr_comparison10}. Specifically, the figure demonstrates that achieving higher statistical power invariably leads to greater privacy leakage, while stronger privacy protection comes at the cost of reduced power. In Figure \ref{fig:rr_comparison10}, the x-axis represents the design parameter $p$, which is the probability of success in the randomization process employed by each RR design. These graphs are based on the hypotheses $ H_0: \pi = 0.2$ versus $H_1: \pi = 0.3$, at a 5\% level of significance, where $\pi$ denotes the proportion of the sensitive group in the population. The sample size is set to 1000. For the unrelated question design, the proportion of the ``unrelated group'' is assumed to be 0.6.

The trade-off between privacy and statistical power in randomized response designs is similar to the well-known balance between Type I and Type II errors in hypothesis testing. If we set a strict limit to avoid Type I errors (false positives), it often becomes harder to avoid Type II errors (false negatives), and vice versa. In our context, enforcing a strong privacy guarantee with a low privacy budget is like imposing a strict bound on the chance of Type I error. This makes it harder to achieve high statistical power, increasing the likelihood of missing true effects in the data. Therefore, finding the best approach for privacy-preserving randomized response designs requires carefully choosing the parameters so that privacy protection does not come at the expense of meaningful or reliable results. The main contribution of this work is the development of a principled optimization framework to design randomized response mechanisms that achieve the highest possible statistical power while satisfying a given differential privacy constraint. Unlike previous approaches that optimize for privacy or for accuracy in isolation, our method simultaneously considers both objectives, enabling the construction of randomized response designs that balance these competing needs in a mathematically rigorous way. We also provide explicit sample size formulas for these optimal designs and offer a free, accessible Shiny web application to support researchers in implementing our approach. This work is the first to provide a unified practical solution for optimizing statistical surveys with formal privacy guarantees under the randomized response paradigm.

The remainder of the article is organized as follows. Section \ref{sec: motivation_example} presents a motivating example. Section \ref{sec:RR-designs} reviews five well-known randomized response designs. Section \ref{sec:DP} introduces differential privacy in the context of RR. Section \ref{sec:optimal_DP} develops optimal differentially private RR designs. Section \ref{sec:simulation} reports simulation studies. Section \ref{sec:data_analysis} demonstrates the real-world applicability of our methods by estimating the proportion of taxpayers who misreport information on their tax returns. Section \ref{sec:shiny_app} describes the accompanying Shiny application. The article concludes with a discussion in Section \ref{sec:discussion}.

\section{MOTIVATING EXAMPLE: INCORRECT INFORMATION ON TAX RETURN}\label{sec: motivation_example}
Our work is motivated by the randomized response study conducted via Amazon Mechanical Turk (AMT), a widely used online crowdsourcing platform \citep{hoglinger2018more}. They were asking sensitive questions, specifically: “Have you ever provided misleading or incorrect information on your tax return?” AMT allows requesters to post tasks, known as Human Intelligence Tasks (HITs), which workers complete in exchange for monetary compensation. Data were collected between November 5, 2013, and December 5, 2013, resulting in 6,505 respondents, all U.S. residents. The sample was fairly balanced by gender (49.9\% male, 50.1\% female), skewed young (with the majority under age 35), and was well educated (88.2\% with at least some college education). Participants were randomly assigned to one of four experimental conditions. Our work focuses on two of these conditions: Direct Questioning (DQ) and Forced Response Design (FRD). The probabilities of the random assignment were \( \frac{1}{8} \) for the DQ condition and \( \frac{2}{8} \) for the FRD condition (the rest of the participants in the other two experimental conditions). A total of 810 respondents were assigned to the DQ condition and 1607 to the FRD condition (see Codebook of ``senstec'', Section A.4, page 95, S1 Documentation of Supporting Information \citet{hoglinger2018more}).  We use the responses from the Direct Questioning condition as reference data against which we compare estimates obtained using the Forced Response Design. Of the assigned participants, 809 individuals in the DQ group and 1602 in the FRD group responded to the sensitive question. 

The Forced Response study design discussed above is based solely on power considerations, without explicit incorporation of privacy budget constraints. In this design, the probabilities of observing outcomes 1 and 6 are taken as $p_1 = \frac{1}{12}$ and $p_2 = \frac{2}{12}$, respectively (see Figure \ref{fig:forced}). This allocation yields a privacy budget estimate of 2.30, which is relatively high and suggests a considerable risk of privacy leakage. A key challenge in this context is that the design does not balance the dual requirements of the collected/generated data, ensuring sufficient statistical power for reliable inference (utility) while simultaneously maintaining adequate privacy protection for sensitive responses. The nontriviality of the problem lies in the inherent trade-off between privacy and utility since increasing privacy protection typically decreases statistical power and vice versa. Moreover, the analysis is further complicated because the privacy budget is a nonlinear function of the spinner's probabilities, $p_1$ and $p_2$. This makes it difficult to analytically identify optimal study designs that jointly maximize utility and minimize privacy risk. Our objective is to rigorously assess whether the Forced Response design could be improved by explicitly accounting for the privacy-utility trade-off, thereby achieving a more efficient and practical balance between accurate inference and robust privacy protection.


\section{RANDOMIZED RESPONSE DESIGNS }\label{sec:RR-designs}
Randomized response designs have been developed to mitigate issues related to misreporting and nonresponse due to privacy concerns, thereby ensuring reliable data collection in surveys involving sensitive questions. Over the years, various designs have been proposed in the literature, each employing distinct mechanisms to introduce randomness while preserving statistical validity. In the following, we summarize the widely used, well-established randomized response designs in chronological order of their introduction in the literature, highlighting the advantages each design offers over its predecessors. For each design, we describe the procedure and discuss how to estimate the population proportion and corresponding variance using the design parameters (e.g., the success probability of spinners) and the sample proportion of ``Yes'' responses obtained from the collected randomized response data. We also provide the exact expressions for the statistical power of each of these designs.

\textbf{Warner's Randomized Response Design:}
Suppose that in the population (see Figure \ref{fig:warner}), individuals belong to one of two distinct groups, such as Group A (sensitive group) or Group B, where B = $A^C$. The objective is to estimate $\pi$, the proportion of Group A in the population, while preserving the privacy of the respondent. A widely used method for achieving this while protecting respondents' privacy is the randomized response design. The first randomized response design was introduced by \cite{warner1965randomized} and is known in the literature as Warner’s randomized response design. This design has been applied in various studies to investigate sensitive topics. For example, \cite{gingerich2010understanding} used Warner's randomized response design to assess corruption levels among public bureaucrats in Bolivia, Brazil, and Chile. In this design, a simple random sample of $n$ individuals is drawn with replacement, and each participant is interviewed. Before interviews, each interviewer receives an identical spinner (such as a coin) that lands on ``A'' with probability $p(\neq \frac{1}{2})$ and on ``B'' with probability $1-p$. During the interview, the interviewer (respondent) privately spins the spinner and reports whether the outcome matches their actual group (see Figure \ref{fig:warner}). If the letter on the spinner corresponds to their group affiliation, they answer ``Yes''; otherwise, they respond ``No''. Importantly, the respondent does not disclose the letter that appeared on the spinner, adding a layer of privacy. This design ensures that individual responses remain confidential while allowing researchers to make statistically valid inferences about the population proportion. Let $\pi$ denote the unknown proportion of individuals in Group A in the population. For each respondent, $i \in \{1, 2, ..., n\}$, let, $X_i = 1$ if the $i^{th}$ respondent reports ``Yes'' and $X_i = 0$ if the $i^{th}$ respondent reports ``No''. Then, $P(X_i = 1) = \pi p + (1 - \pi) (1 - p)$, and $P(X_i = 0) = (1 - \pi) p + \pi (1 - p)$. Hence, an unbiased estimator of $\pi$ is $\hat{\pi} = \frac{p - 1}{2p - 1} + \frac{\hat{\pi}_s}{2p - 1}$, where, $\hat{\pi}_s$ is the sample proportion of ``Yes'' responses and $V(\hat{\pi}) = \frac{1}{n} [\frac{1}{16(p - \frac{1}{2})^2} - (\pi - \frac{1}{2})^2].$ 

\textbf{Unrelated Question Randomized Response Design:} Warner’s randomized response design allows respondents to answer either the sensitive question or its negation based on a randomizing device. This method helps in estimating the prevalence of sensitive characteristics without revealing individual responses. However, Warner’s model has limitations, particularly regarding respondent suspicion and compliance. Some individuals may still feel uncomfortable, fearing their response indirectly discloses sensitive information. To overcome this concern, the Unrelated Question Randomized Response (UQRR) Design was proposed as an alternative (see Figure \ref{fig:unrelated}). Introduced by \cite{simmons1967unrelated} and \cite{greenberg1969unrelated}, UQRR incorporates an unrelated, non-sensitive question alongside the sensitive one. Respondents are instructed to answer one of these questions based on the outcome of a randomizing device. Since the unrelated question is harmless and independent of the sensitive characteristic, this approach reduces suspicion and improves response accuracy. For example, in UQRR design, a respondent uses a randomizing device to answer either a sensitive question such as ``Have you ever used illicit drugs?'' or an unrelated, non-sensitive question such as ``Is your birthday in January, February, or March?'', with the unrelated question being chosen with a known probability (approximately $0.25$ in this case). The respondent reports only ``Yes'' or ''No,'' without revealing which question was answered, thereby preserving privacy. One of the earliest applications of UQRR was in estimating induced abortion rates in urban North Carolina \cite{abernathy1970estimates}. Similar studies have used this method in Mexico \citep{lara2004measuring, lara2006measure} and Turkey \citep{tezcan1981prevalence} to assess abortion prevalence.

In this approach (see Figure \ref{fig:unrelated}), rather than directly identifying whether a respondent belongs to the sensitive group (Group A) or its complement (Group B), the respondent provides a ``Yes'' or ``No'' answer to one of two possible questions. One question pertains to Group A, while the other relates to an unrelated, non-sensitive group (Group Y). Statistically, membership in Group Y is independent of Group A. A randomization device, such as a spinner or a deck of cards, determines which question the respondent answers. The objective is to estimate $\pi$, the proportion of the population in Group A. The probability of being assigned the question about Group A is $p > 0$, while the probability of being assigned the question about Group Y is $1-p$. Group Y should be entirely unrelated to the sensitive characteristic and have a known population proportion, denoted as $\pi_Y$. A simple random sample of $n$ individuals is drawn with replacement, and each respondent answers based solely on the randomly assigned question. Let $\pi$ denote the unknown proportion of individuals in Group A in the population. For each respondent, $i \in \{1, 2, ..., n\}$, let, $X_i = 1$ if the $i^{th}$ respondent reports ``Yes'' and $X_i = 0$ if the $i^{th}$ respondent reports ``No''. Then, $P(X_i = 1) = p \pi + (1 - p) \pi_Y$. Hence, an unbiased estimator of $\pi$ is, $\hat{\pi} = \frac{\hat{\pi}_s - (1 - p) \pi_Y}{p}$, where, $\hat{\pi}_s$ is the sample proportion of ``Yes'' responses and $V(\hat{\pi}) = \frac{[p \pi + (1 - p) \pi_Y] - [p \pi + (1 - p) \pi_Y]^2}{n p^2}.$

\textbf{Forced Response Design:} The Forced Response Design (FRD), first introduced by \cite{boruch1971assuring} and later refined by \cite{fox1986randomized}, is a widely used randomized response technique that enhances respondent privacy (see Figure \ref{fig:forced}). The Unrelated Question Randomized Response Design relies on a neutral question to mask responses and requires knowing the true proportion of the population who would answer the unrelated question in a certain way. This information can sometimes be difficult to obtain. In contrast, the FRD eliminates the need for an unrelated question, making the method simpler and more practical to implement. This method follows a randomized response mechanism in which individuals answer truthfully or provide a predetermined response. A common implementation (see Figure \ref{fig:forced}) involves rolling a die: if it lands on $1$, the respondent must answer ``No'' regardless of their true status; if it lands on $6$, they must answer ``Yes''; otherwise, they respond truthfully. If $p_1$ and $p_2$ denote the probabilities of rolling a 1 and a 6, respectively, then the probability of providing a truthful response is $1 - p_1 - p_2$, with $p_1 + p_2 < 1$. This approach ensures confidentiality while allowing researchers to estimate the true proportion of individuals with sensitive characteristics. A simple random sample of $n$ individuals is selected with replacement, and each respondent follows the described randomized response mechanism. Let $\pi$ denote the unknown proportion of individuals in Group A (sensitive group) in the population. For each respondent, $i \in \{1, 2, ..., n\}$, let, $X_i = 1$ if the $i^{th}$ respondent reports ``Yes'' and $X_i = 0$ if the $i^{th}$ respondent reports ``No''. Then, $P(X_i = 1) = p_2 + (1 - p_1 - p_2) \pi$. Hence, an unbiased estimator of $\pi$ is, $\hat{\pi} = \frac{\hat{\pi}_s - p_2}{1 - p_1 - p_2}$, where, $\hat{\pi}_s$ is the sample proportion of ``Yes'' responses and $V(\hat{\pi}) = \frac{[p_2 + (1 - p_1 - p_2) \pi] - [p_2 + (1 - p_1 - p_2) \pi]^2}{n (1 - p_1 - p_2)^2}.$
This design has been used to estimate the prevalence of sensitive issues such as xenophobia and anti-semitism in Germany \citep{krumpal2012estimating}, as well as to assess whether individuals voted for an anti-abortion referendum during the 2011 Mississippi General Election \citep{rosenfeld2016empirical}. In addition, it has been used to study illegal activities, including poaching among South African farmers \citep{st2012identifying} and the use of performance-enhancing drugs among members of fitness centers \citep{stubbe2014prevalence}. These diverse applications demonstrate the utility of FRD in obtaining accurate data while protecting the anonymity of the respondent.

\textbf{Kuk's Randomized Response Design:} A significant limitation of traditional randomized response designs is that some respondents must answer ``Yes'' or ``No'' to a sensitive question, which can lead to discomfort, skepticism, and reduced participation. Ensuring that respondents feel genuinely protected by the randomization process remains a challenge. \cite{warner1986omitted} observed that interviewees often behave as if the device merely determines whether they must disclose sensitive information. Furthermore, \cite{warner1986omitted} and \cite{fox1986randomized} noted that respondents without sensitive characteristics can overemphasize their innocence, introducing bias. To address these concerns, \cite{kuk1990asking} proposed (see Figure \ref{fig:kuk}) an alternative design that eliminates the need for direct responses, thereby improving respondent cooperation. In this approach, respondents generate two binary outcomes independently using Bernoulli distributions with known probabilities $p_1 > 0$ and $p_2 > 0$. If they belong to Group A, they report the first outcome; otherwise, they report the second. A practical way to implement this is by using two separate decks of cards, each consisting of only red and black cards, where the proportion of red cards in each deck corresponds to $p_1$ and $p_2$, respectively, with $p_1 \neq p_2$. After shuffling, the respondent randomly draws one card from each deck and reports an outcome based on the drawn colors. Since respondents never directly answer the sensitive question, this method minimizes discomfort and promotes more truthful participation. A simple random sample of $n$ individuals is chosen with replacement, and each follows the specified randomized response procedure. Let $\pi$ denote the unknown proportion of individuals in the population who belong to the sensitive group, Group A. For each respondent, $i \in \{1, 2, ..., n\}$, let, $X_i = 1$ if the $i^{th}$ respondent reports ``red'' and $X_i = 0$ if the $i^{th}$ respondent reports ``black''. Then, $P(X_i = 1) = p_1 \pi + p_2 (1 - \pi)$. Hence, an unbiased estimator of $\pi$ is, $\hat{\pi} = \frac{\hat{\pi}_s - p_2}{p_1 - p_2}$, where, $\hat{\pi}_s$ is the sample proportion of ``red'' responses and $V(\hat{\pi}) = \frac{[p_1 \pi + p_2 (1 - \pi)] - [p_1 \pi + p_2 (1 - \pi)]^2}{n (p_1 - p_2)^2}$.
\cite{edgell1982validity} used a forced response method to study the experiences of college students related to homosexuality. However, their findings showed that $25\%$ of the respondents instructed to answer ``Yes'' did not comply, indicating that the method may still cause discomfort. Recognizing this issue, \cite{vanderheijden1996some} suggested that Kuk's design could be more effective in reducing the discomfort of the respondent and improving the reliability of the data.

\textbf{A Two-Step Randomized Response Design:} Figure \ref{fig:twostep} illustrates a randomized response design that has recently gained popularity, particularly among computer science researchers \citep{chang2023privacy, dwork2014algorithmic}. The population is divided into two groups: a sensitive group, Group A, and its complementary group, Group B (see Figure \ref{fig:twostep}). The objective is to estimate the proportion of individuals in Group A within the population. Each respondent flips a specially designed coin that lands on heads with probability $p > 0$. If the first flip results in heads, the respondent answers truthfully. If it results in tails, they flip a second, identical coin with the same probability of heads, $p$. In this case, they respond ``Yes'' if the second flip is heads and ``No'' if it is tails. This approach ensures that individual responses remain confidential while enabling accurate estimation of population characteristics. A simple random sample of $n$ individuals is drawn with replacement from the population, and each selected individual follows this randomized response procedure. Let $\pi$ denote the unknown proportion of individuals in Group A in the population. For each respondent, $i \in \{1, 2, ..., n\}$, let, $X_i = 1$ if the $i^{th}$ respondent reports ``Yes'' and $X_i = 0$ if the $i^{th}$ respondent reports ``No''. Then, $P(X_i = 1) = p \pi + p (1 - p)$. Hence, an unbiased estimator of $\pi$ is, $\hat{\pi} = \frac{\hat{\pi}_s}{p} - (1 - p)$, where, $\hat{\pi}_s$ is the sample proportion of ``Yes'' responses and $V(\hat{\pi}) = \frac{[p \pi + p (1 - p)] - [p \pi + p (1 - p)]^2}{np^2}$. However, this model shares the same limitation as Warner’s RR design, as respondents, particularly those who answer ``Yes'', may still experience suspicion or discomfort. They may fear that their answers could indirectly disclose sensitive information, which affects their willingness to comply fully with the procedure.

\subsection{HYPOTHESIS TESTING}\label{hypothesis_testing}
The objective of hypothesis testing in randomized response designs is to formally assess whether the proportion, $\pi$, of individuals in the population who belong to a sensitive group equals a specific value of interest. This assessment is important not only for drawing valid statistical inferences about sensitive characteristics but also for determining the appropriate sample size required to achieve a desired level of statistical power when designing a new study. Suppose we wish to test whether the proportion $\pi$ equals a predefined value $\pi_0$. The hypothesis testing problem is stated as $H_0: \pi = \pi_0 \quad \text{vs.} \quad H_1: \pi \neq \pi_0$ at a significance level $\alpha$. Let $T$ be an estimator of $\pi$ derived from the corresponding randomized response design. We use the Wald test approach assuming that $T$ follows an asymptotically normal distribution. The null hypothesis is rejected if, $T > \pi_0 + z_{\frac{\alpha}{2}} \sqrt{V_{\pi_0}(T)}$, or, $T < \pi_0 - z_{\frac{\alpha}{2}} \sqrt{V_{\pi_0}(T)}$, where \( z_{\frac{\alpha}{2}} \) is the upper \( \frac{\alpha}{2} \)-quantile of the standard normal distribution. Therefore, the power of the test is given by, 
\begin{align*}
P_{\pi_1}(\text{Rejecting } H_0) 
    &= P_{\pi_1}\left( T < \pi_0 - z_{\frac{\alpha}{2}} \sqrt{V_{\pi_0}(T)} \right) 
       + P_{\pi_1}\left( T > \pi_0 + z_{\frac{\alpha}{2}} \sqrt{V_{\pi_0}(T)} \right) \\
    &= \Phi(d_1) + 1 - \Phi(d_2),
\end{align*}
where $d_1 = \frac{\pi_0 - \pi_1 - z_{\frac{\alpha}{2}} \sqrt{V_{\pi_0}(T)}}{\sqrt{V_{\pi_1}(T)}}$, $d_2 = \frac{\pi_0 - \pi_1 + z_{\frac{\alpha}{2}} \sqrt{V_{\pi_0}(T)}}{\sqrt{V_{\pi_1}(T)}}$, $\Phi$ is the cumulative distribution function (CDF) of standard normal distribution, and $\pi = \pi_1$ under $H_1$. Table \ref{Table:RR-power} summarizes the statistical power corresponding to each of the five randomized response designs considered in this work.

\begin{table*}[t]
\centering
\caption{Power of different randomized response designs.}
\label{Table:RR-power}

\renewcommand{\arraystretch}{2.5}
\tabcolsep=8pt

\begin{tabular}{|
>{\centering\arraybackslash}m{2cm}|
>{\centering\arraybackslash}m{0.8\textwidth}|
}
\hline
\textbf{Design} & \textbf{Power} \\
\hline

Warner &
\small$\displaystyle
\begin{aligned}
1 + \Phi\!\left(
\frac{2\sqrt{n}(\pi_{0}-\pi_{1})(2p-1) - z_{\alpha/2}\sqrt{1-(2\pi_{0}-1)^2(2p-1)^2}}
{\sqrt{1-(2\pi_{1}-1)^2(2p-1)^2}}
\right) \\
-\; \Phi\!\left(
\frac{2\sqrt{n}(\pi_{0}-\pi_{1})(2p-1) + z_{\alpha/2}\sqrt{1-(2\pi_{0}-1)^2(2p-1)^2}}
{\sqrt{1-(2\pi_{1}-1)^2(2p-1)^2}}
\right)
\end{aligned}
$ \\
\hline

Unrelated Question &
\small$\displaystyle
\begin{aligned}
1 + \Phi\!\left(
\frac{
p\sqrt{n}\,(\pi_{0}-\pi_{1})
- z_{\alpha/2}\sqrt{[p\pi_{0} + (1-p)\pi_{Y}] - [p\pi_{0} + (1-p)\pi_{Y}]^{2}}
}{
\sqrt{[p\pi_{1} + (1-p)\pi_{Y}] - [p\pi_{1} + (1-p)\pi_{Y}]^{2}}
}
\right)
\\
-\,
\Phi\!\left(
\frac{
p\sqrt{n}\,(\pi_{0}-\pi_{1})
+ z_{\alpha/2}\sqrt{[p\pi_{0} + (1-p)\pi_{Y}] - [p\pi_{0} + (1-p)\pi_{Y}]^{2}}
}{
\sqrt{[p\pi_{1} + (1-p)\pi_{Y}] - [p\pi_{1} + (1-p)\pi_{Y}]^{2}}
}
\right)
\end{aligned}
$ \\
\hline

Forced Response &
\scalebox{0.75}{$\displaystyle
\begin{aligned}
1 + \Phi\!\left(
\frac{
\sqrt{n}\,(1-p_{1}-p_{2})(\pi_{0}-\pi_{1})
- z_{\alpha/2}\sqrt{[p_{2}+(1-p_{1}-p_{2})\pi_{0}] - [p_{2}+(1-p_{1}-p_{2})\pi_{0}]^{2}}
}{
\sqrt{[p_{2}+(1-p_{1}-p_{2})\pi_{1}] - [p_{2}+(1-p_{1}-p_{2})\pi_{1}]^{2}}
}
\right)
\\
-\,
\Phi\!\left(
\frac{
\sqrt{n}\,(1-p_{1}-p_{2})(\pi_{0}-\pi_{1})
+ z_{\alpha/2}\sqrt{[p_{2}+(1-p_{1}-p_{2})\pi_{0}] - [p_{2}+(1-p_{1}-p_{2})\pi_{0}]^{2}}
}{
\sqrt{[p_{2}+(1-p_{1}-p_{2})\pi_{1}] - [p_{2}+(1-p_{1}-p_{2})\pi_{1}]^{2}}
}
\right)
\end{aligned}
$} \\
\hline

Kuk &
\scalebox{0.80}{$\displaystyle
\begin{aligned}
1 + \Phi\!\left(
\frac{
\sqrt{n}\,(p_{1}-p_{2})(\pi_{0}-\pi_{1})
- z_{\alpha/2}\sqrt{[p_{1}\pi_{0}+p_{2}(1-\pi_{0})] - [p_{1}\pi_{0}+p_{2}(1-\pi_{0})]^{2}}
}{
\sqrt{[p_{1}\pi_{1}+p_{2}(1-\pi_{1})] - [p_{1}\pi_{1}+p_{2}(1-\pi_{1})]^{2}}
}
\right)\\
-\,
\Phi\!\left(
\frac{
\sqrt{n}\,(p_{1}-p_{2})(\pi_{0}-\pi_{1})
+ z_{\alpha/2}\sqrt{[p_{1}\pi_{0}+p_{2}(1-\pi_{0})] - [p_{1}\pi_{0}+p_{2}(1-\pi_{0})]^{2}}
}{
\sqrt{[p_{1}\pi_{1}+p_{2}(1-\pi_{1})] - [p_{1}\pi_{1}+p_{2}(1-\pi_{1})]^{2}}
}
\right)
\end{aligned}
$} \\
\hline

Two-Step &
\small$\displaystyle
\begin{aligned}
1 + \Phi\!\left(
\frac{
\sqrt{n}\,p(\pi_{0}-\pi_{1}) 
- z_{\alpha/2}\sqrt{[p\pi_{0}+p(1-p)] - [p\pi_{0}+p(1-p)]^{2}}
}{
\sqrt{[p\pi_{1}+p(1-p)] - [p\pi_{1}+p(1-p)]^{2}}
}
\right)
\\
-
\Phi\!\left(
\frac{
\sqrt{n}\,p(\pi_{0}-\pi_{1}) 
+ z_{\alpha/2}\sqrt{[p\pi_{0}+p(1-p)] - [p\pi_{0}+p(1-p)]^{2}}
}{
\sqrt{[p\pi_{1}+p(1-p)] - [p\pi_{1}+p(1-p)]^{2}}
}
\right)
\end{aligned}
$ \\
\hline

\end{tabular}
\end{table*}

\section{DIFFERENTIAL PRIVACY}\label{sec:DP}
In this section, we introduce the concept of differential privacy and examine its relationship to randomized response designs. Differential privacy has emerged as a rigorous framework for quantifying and ensuring the privacy of individuals in statistical analyses \citep{dwork2006differential, dwork2014algorithmic}. Unlike traditional methods, such as randomized response, which provide privacy through probabilistic masking at the individual level, differential privacy offers formal guarantees that limit the influence of any single respondent’s data on the overall output. We review the formal definition of differential privacy and discuss its interpretation in the context of randomized response. We also highlight recent developments that leverage differential privacy principles to enhance privacy protection in survey sampling. To rigorously define differential privacy, we first introduce some foundational concepts regarding randomized algorithms and databases.

\begin{definition}[Probability Simplex \citep{dwork2014algorithmic}] Suppose $B$ is a discrete set. Then the probability simplex over $B$, denoted $\Delta(B)$, is defined by
\[
\Delta(B) = \left\{ x \in \mathbb{R}^{|B|} \;:\; x_i \ge 0 \ \forall i \text{ and } \sum_{i=1}^{|B|} x_i = 1 \right\}.
\]

\end{definition}

\begin{definition}[Randomized Algorithm \citep{dwork2014algorithmic}] A randomized algorithm $\mathcal{M}$ operates on a domain $A$ and produces outputs in a discrete range $B$. It is characterized by a mapping $f: A \to \Delta(B)$. For any input $a \in A$, the algorithm $\mathcal{M}$ produces an output $\mathcal{M}(a) = b$ with probability $(f(a))_b$, for each $b \in B$. 
\end{definition}

Let $D$ be a database consisting of a collection of records, each drawn from a universal set $\mathcal{X}$. That is, $D \in \mathcal{X}^n$ for a dataset of size $n$. The central concept of differential privacy is defined as follows:

\begin{definition}[Differential Privacy \citep{dwork2006differential, wang2016using}] A randomized algorithm $\mathcal{M}$ provides $\varepsilon$-differential privacy if, for any two databases $D$ and $D'$ differing in at most one record, and for any subset $S$ of possible outputs of the output range of $\mathcal{M}$, the following holds: 
$$P[\mathcal{M}(D) \in S] \leq e^{\varepsilon} P[\mathcal{M}(D') \in S].$$
\end{definition}

Differential privacy guarantees that the inclusion or exclusion of any single individual's data can alter the probability of any particular output by at most a multiplicative factor of $e^{\varepsilon}$. Formally, for any possible output, the ratio between the probabilities of that output under two neighboring datasets (differing by one individual or record) is bounded above by $e^{\varepsilon}$. As a result, the output distribution remains statistically similar regardless of whether any particular individual is represented in the dataset, thus safeguarding personal privacy. This property ensures that it is infeasible to confidently infer whether an individual participated in the dataset or what their true data was, based solely on the observed output. For instance, in the randomized response mechanism, the probability of observing a response such as ``Yes'' is nearly indistinguishable whether or not a single individual's true answer is included, as any change in probability is strictly controlled and bounded by $e^{\varepsilon}$. Thus, differential privacy rigorously formalizes privacy protection by ensuring that the influence of any one individual's data on the analysis outcome is tightly limited.

The parameter $\varepsilon$ is known as the differential privacy budget, which determines how much information a mechanism is allowed to reveal. A higher $\varepsilon$ means weaker privacy protection, as it allows more information leakage, whereas a lower $\varepsilon$ ensures stricter privacy by limiting the distinguishability of the outputs. In practical applications, $\varepsilon$ is usually kept small to maintain strong privacy guarantees. For statistical analyses such as estimation of means or frequencies, $\varepsilon$ is typically set between 0.001 and 1, ensuring minimal disclosure of information \citep{chang2023privacy}.

Although differential privacy provides strong privacy guarantees in centralized settings where a trusted curator collects and analyzes the data, many real-world scenarios involve data collection in untrusted or potentially adversarial environments. In the standard DP framework, privacy is ensured by bounding the ratio of output probabilities for any two neighboring datasets $D$ and $D'$ that differ by a single individual's record, but this guarantee relies on the assumption that the curator can be trusted to apply the privacy-preserving mechanism correctly. However, in situations where such trust cannot be assumed, stronger privacy protection is necessary, specifically guarantees that operate at the individual level before any raw data leaves a user or their device. This need has motivated the development of \emph{Local Differential Privacy (LDP)}, a variant of differential privacy in which each participant independently perturbs their own data through a randomized algorithm prior to sharing. Under LDP, the privacy of each user is protected against any external observer, including the data collector, as the mechanism ensures that the probability distributions over outputs remain similar whether the individual's true data is $x$ or some alternative $x'$. Thus, LDP extends the core principles of differential privacy to settings where the change from $D$ to $D'$, capturing the effect of a single individual's data, must be safeguarded even before data aggregation begins.

\begin{definition}[Local Differential Privacy \citep{chang2023privacy}] A randomized algorithm $\mathcal{M}$ is said to satisfy $\varepsilon$-local differential privacy if, for any pair of possible inputs $x$ and $x'$, and for any possible output $y$, the following inequality holds:
$$P[\mathcal{M}(x) = y] \leq e^{\varepsilon}  P[\mathcal{M}(x') = y].$$
\end{definition}
This ensures that the output distribution produced by $\mathcal{M}$ is similar whether it operates on $x$ or $x'$, thereby providing strong privacy guarantees depending on the value of $\varepsilon$ for each individual's data before it is collected or aggregated. Local differential privacy (LDP) is particularly relevant for privacy-preserving data collection methods, such as randomized response, in which participants perturb their responses locally before sharing. Since randomized response mechanisms inherently operate at the individual level, LDP is the appropriate privacy measure for analyzing these designs. 


\begin{theorem}
The Warner randomized response mechanism satisfies $\varepsilon$-differential privacy with
\[
\varepsilon = \max\left\{ \ln \frac{p}{1-p},\ \ln \frac{1-p}{p} \right\}, \quad p \neq \tfrac{1}{2}.
\]
\end{theorem}

\begin{proof}

In Warner's randomized response design, if a respondent belongs to Group A (or Group B) and the spinner points to A (or B), the respondent answers ``Yes''. The probability that the spinner points to A is $p$ and the probability that it points to B is $1-p$. \\Let, for the $i^{th}$ individual respondent, where $i \in \{1, 2, ..., n\}$,

\small\[
X_i =
\begin{cases}
1, & \text{if } i^{th}\text{ respondent reports ``Yes''}, \\
0, & \text{if } i^{th}\text{ respondent reports ``No''}.
\end{cases}
\quad
Z_i =
\begin{cases}
1, & \text{if } i^{th}\text{ respondent belongs to group A}, \\
0, & \text{if } i^{th}\text{ respondent belongs to group B}.
\end{cases}
\]
\\Thus, the conditional distribution of $X_i$ given $Z_i$ is
\[
P(X_i = 1 \mid Z_i = 1) = p, \quad
P(X_i = 0 \mid Z_i = 1) = 1-p,
\]
\[
P(X_i = 1 \mid Z_i = 0) = 1-p, \quad
P(X_i = 0 \mid Z_i = 0) = p.
\]
\\To show that Warner’s randomized response design satisfies differential privacy, consider any two neighboring inputs $Z_i = 0$ and $Z_i = 1$, and any possible output $X_i \in \{0,1\}$. Then 
\[
\frac{P(X_i = 1 \mid Z_i = 1)}{P(X_i = 1 \mid Z_i = 0)} = \frac{p}{1-p}, 
\quad
\frac{P(X_i = 0 \mid Z_i = 1)}{P(X_i = 0 \mid Z_i = 0)} = \frac{1-p}{p}.
\]
Similarly,
\[
\frac{P(X_i = 1 \mid Z_i = 0)}{P(X_i = 1 \mid Z_i = 1)} = \frac{1-p}{p}, 
\quad
\frac{P(X_i = 0 \mid Z_i = 0)}{P(X_i = 0 \mid Z_i = 1)} = \frac{p}{1-p}.
\]
\\Therefore, for all possible output

\vspace{0.1cm}
\normalsize$\max \left\{
\frac{P(X_i = 1 \mid Z_i = 1)}{P(X_i = 1 \mid Z_i = 0)},\ 
\frac{P(X_i = 0 \mid Z_i = 1)}{P(X_i = 0 \mid Z_i = 0)},\ 
\frac{P(X_i = 1 \mid Z_i = 0)}{P(X_i = 1 \mid Z_i = 1)},\ 
\frac{P(X_i = 0 \mid Z_i = 0)}{P(X_i = 0 \mid Z_i = 1)} 
\right\}
= \max\left\{ \frac{p}{1-p},\ \frac{1-p}{p} \right\}$.
\vspace{0.1cm}
\\ Therefore, the design satisfies $\varepsilon$-differential privacy with
\[
\varepsilon = \max\left\{ \ln\frac{p}{1-p},\ \ln\frac{1-p}{p} \right\}.
\]
\end{proof}

The above theorem establishes that Warner’s randomized response design satisfies differential privacy and provides an explicit expression for the corresponding privacy budget. Using analogous arguments, the differential privacy budgets for the remaining randomized response designs can be derived. For completeness, Table \ref{Table: RR-dp} summarizes the differential privacy budgets for all five designs considered in this paper.

\begin{table}[h!]
    \renewcommand{\arraystretch}{2} 
    \centering
    \caption{Differential privacy budget formulas for the considered randomized response designs. The $\ln$ denotes the logarithm with base $e$.}
    \begin{adjustbox}{center}
    \begin{tabular}{|p{7cm}|p{13cm}|}
    \hline
    \multicolumn{1}{|c|}{\textbf{Design}} & \multicolumn{1}{c|}{\textbf{Differential Privacy Budget ($\varepsilon$)}} \\
    \hline
    \multicolumn{1}{|c|}{Warner} & \multicolumn{1}{|c|}{$\max\{\ln\frac{p}{1-p}, \ln\frac{1 - p}{p}\}$}\\
    
    \hline
    \multicolumn{1}{|c|}{Unrelated Question} & \multicolumn{1}{|c|}{$\max\{\ln \frac{p + (1 - p) \pi_Y}{\pi_Y (1-p)}, \ln \frac{p + (1 - \pi_Y)(1 - p)}{(1 - \pi_Y)(1 - p)}\}$}\\
    \hline
    \multicolumn{1}{|c|}{Forced Response} & \multicolumn{1}{|c|}{$\max \{\ln \frac{1 - p_1}{p_2}, \ln \frac{1 - p_2}{p_1}\}$}\\
    \hline
    \multicolumn{1}{|c|}{Kuk} & \multicolumn{1}{|c|}{$\max\{\ln \frac{p_1}{p_2}, \ln \frac{p_2}{p_1}, \ln \frac{1 - p_1}{1 - p_2}, \ln \frac{1 - p_2}{1 - p_1}\}$}\\
    \hline
    \multicolumn{1}{|c|}{Two-Step} & \multicolumn{1}{|c|}{$\max\{\ln \frac{p + p(1 - p)}{p(1 - p)}, \ln \frac{p + (1 - p)(1 - p)}{(1 - p)(1 - p)}\}$}\\
    \hline
    \end{tabular}
    \label{Table: RR-dp}
    \end{adjustbox}
\end{table}

In Table~\ref{Table:RR-power}, we observe that the statistical power of all the randomized response designs under discussion depends on the sample size ($n$), the hypothesized values of the population parameter $\pi$ under $H_0$ and $H_1$, and the design parameter(s) ($p$). In contrast, Table~\ref{Table: RR-dp} shows that the differential privacy (DP) budget depends solely on the design parameter(s) ($p$) and is independent of the sample size (UQRR also depends on known $\pi_Y$). Therefore, if an investigator wishes to conduct a randomized response study, the minimum required sample size cannot be determined from the differential privacy budget alone. In other words, the DP budget $\varepsilon$ is invariant with respect to the sample size $n$. Similarly, determining the sample size for a randomized response study based solely on statistical power may result in inadequate privacy protection and potential privacy leakage. This necessitates the consideration of both statistical power and differential privacy together, so that issues related to both accuracy and privacy can be addressed in a unified framework.


\section{OPTIMAL DIFFERENTIALLY PRIVATE RANDOMIZED RESPONSE DESIGNS}\label{sec:optimal_DP}
In previous sections, we discussed the statistical power and differential privacy budget associated with various randomized response designs. High statistical power ensures an accurate estimation of the proportion of the sensitive group in the population ($\pi$), while a lower differential privacy budget ($\varepsilon$) provides a stronger privacy guarantee for participants' data when using RR designs. Therefore, both statistical power and the privacy budget are crucial criteria in the design of a randomized response study. 

However, as highlighted in the Introduction and illustrated in Figure \ref{fig:rr_comparison10}, there exists an inherent trade-off between power and the DP budget; achieving higher statistical power typically requires a higher (i.e., less strict) DP budget, and vice versa. As mentioned in the Introduction, this trade-off is analogous to the balance between Type I and Type II errors in hypothesis testing, where the common practice is to fix the Type I error rate and then minimize the Type II error. Drawing on this analogy, our approach is to restrict the privacy budget first and then maximize the statistical power with respect to other study parameters. This privacy-first approach is justified because the DP budget is determined solely by the randomization parameter $p$ (or the vector $p$ in designs such as Forced Response and Kuk’s Designs). Investigators can specify the desired level of privacy protection and set $\varepsilon$ accordingly, informed by ethical, legal, or institutional requirements. In contrast, statistical power is influenced by multiple factors, including the null and alternative proportions of the sensitive group ($\pi_0$, $\pi_1$), the randomization parameter $p$, and the sample size $n$. The values of $\pi_0$ and $\pi_1$ are set from the research hypothesis that addresses the primary research question. Therefore, by first selecting an appropriate privacy budget, we ensure compliance with privacy requirements, and then we optimize the study design to achieve the highest possible power within this constraint. We assume that $\pi_0$ and $\pi_1$ (or equivalently, their difference $\pi_1 - \pi_0$) are known or specified by the research hypothesis. The objective is to maximize the statistical power of the randomized response test stated in Section \ref{hypothesis_testing}, subject to a constraint on the differential privacy budget. Therefore, to obtain the optimal values of $n$ and $p$, we first define the feasible set of privacy budget of $p$  for a given privacy budget $c > 0$,
\begin{eqnarray}
\mathbb{P}(c) = \{\, p : \varepsilon(p) \le c \,\},
\label{eq:privacy_budget-set}
\end{eqnarray}
where $p$ may be either a scalar or a vector, depending on the design, and $\varepsilon(p)$ denotes privacy budget as a function of $p$. For each $p \in \mathbb{P}(c)$, the minimum sample size is
\begin{eqnarray}
n(p) = \min \left\{ n \in \mathbb{N} : \Phi\!\big(d_1(n,p)\big) + 1 - \Phi\!\big(d_2(n,p)\big) \ge 1 - \beta \right\},
\label{eq:sample_size_n-set}
\end{eqnarray}
where $\beta \in (0,1)$ denotes the Type II error rate (so $1 - \beta$ is the target power), $\mathbb{N} = \{1,2,3,\ldots\}$ is the set of positive integers, and $\Phi$ is the standard normal cumulative distribution function. The minimum required sample size and an optimal design parameter are then
\begin{eqnarray}
n^* = \min_{p \in \mathbb{P}(c)} n(p), \mbox{ and } p^* = \arg\min_{p \in \mathbb{P}(c)} n(p).
\label{eq:optimum_n_p}    
\end{eqnarray}

In this formulation, for each feasible $p$ satisfying $\varepsilon(p) \leq c$, we identify the minimum sample size $n$ required to achieve the desired power. The optimal solution $(n^*, p^*)$ is the pair that yields the smallest such $n$. The optimization problem described in (\ref{eq:optimum_n_p}) can be solved by the following binary search Algorithm \ref{Algo:joint_optimization}.

\begin{algorithm}[ht]
\caption{Joint Optimization for $(n^*, p^*)$}
\label{Algo:joint_optimization}
\begin{algorithmic}[1]

\Require Desired power $1-\beta$, privacy budget $c$, maximum sample size $N_{\max}$
\Ensure Minimum sample size $n^*$ and corresponding $p^*$

\State $n^* \gets \text{null}$
\State $p^* \gets \text{null}$

\For{each $p$ such that $\varepsilon(p) \leq c$}
    \State $l \gets 1$
    \State $r \gets N_{\max}$
    \State $n_p \gets \text{null}$

    \While{$l \leq r$}
        \State $n \gets \lfloor (l + r)/2 \rfloor$
        \State $P \gets \Phi(d_1(n,p)) + 1 - \Phi(d_2(n,p))$

        \If{$P \geq 1 - \beta$}
            \State $n_p \gets n$
            \State $r \gets n - 1$ \Comment{search lower}
        \Else
            \State $l \gets n + 1$ \Comment{search higher}
        \EndIf

    \EndWhile

    \If{$n_p \neq \text{null}$}
        \If{$n^* = \text{null}$ \textbf{or} $n_p < n^*$}
            \State $n^* \gets n_p$
            \State $p^* \gets p$
        \EndIf
    \EndIf

\EndFor

\Return $n^*,\, p^*$

\end{algorithmic}
\end{algorithm}


In many practical randomized response surveys, the sample size $n$ is often determined in advance due to limitations imposed by budget, logistics, or regulatory requirements. For instance, the number of participants may be restricted by available resources, administrative decisions, or ethical review board approvals. Consequently, $n=n_0$ is treated as a fixed parameter in the subsequent design optimization, and attention is focused on identifying the optimal design parameter $p$ that maximizes the power and meets feasibility constraints. Given a privacy budget $c>0$ and the feasible set of privacy budget $\mathbb{P}(c)$ defined in (\ref{eq:privacy_budget-set}), the optimal design parameter is
\begin{eqnarray}
p^* = \arg\max_{p \in \mathbb{P}(c)} \left\{ \Phi\big(d_1(n_0, p)\big) + 1 - \Phi\big(d_2(n_0, p)\big) \right\}.
 \label{eq:single_para_p_optimization}
\end{eqnarray}

In (\ref{eq:single_para_p_optimization}), for a fixed sample size $n = n_0$, it may occur that there is no value of $p$ that satisfies the privacy constraint $\varepsilon(p) \leq c$ and achieves a given statistical power at least $1 - \beta$. This situation arises when the prescribed privacy budget $c$ is too stringent relative to the available sample size $n_0$, making it impossible to select a $p$ that yields sufficient statistical power. In such cases, one may need to relax the privacy constraint, increase the sample size, or accept a lower power level.

In some settings, the sample size is fixed, and the objective is to maintain power at least $1-\beta$ while minimizing the privacy budget. In that case, the optimal design parameter(s) can be obtained by first identifying the set of values of parameters that achieve the target power for the given sample size $n_0$, and then selecting from that set the parameter(s) that minimize the privacy budget. Define the feasible set of $p$ for a given desired power of $1-\beta$ and fixed sample size $n_0$ as
\begin{eqnarray}
\mathbb{P}(n_0,\beta)=\{p: \Phi\big(d_1(n_0, p)\big) + 1 - \Phi\big(d_2(n_0, p)\big) \geq 1-\beta\}.  
\label{eq:single_para_p_optimization_fixed_power}
\end{eqnarray}
Then the optimal design parameter is
\begin{equation}
p^*=\arg\min_{p \in\mathbb{P}(n_0,\beta)}\ \varepsilon(p).
 \label{eq:single_para_p_optimization_min_privacy}
\end{equation}
If $\mathbb{P}(n_0,\beta)$ is empty, one must increase $n_0$ or relax the power requirement.

\begin{theorem}
\label{thm:optimal_boundary_warner}
Consider Warner’s randomized response design with design parameter $p\in(0,1)\setminus\left\{\tfrac12\right\}$, differential privacy budget $\varepsilon(p)$, and power function $\mathcal{P}(p)$ for a level-$\alpha$ test of $H_0: \pi = \pi_0$ vs. $H_1: \pi = \pi_1$, where $\pi_0 \ne \pi_1$, with sample size $n$. Then,

\begin{enumerate}
\item The power function $\mathcal{P}(p)$ is symmetric about $p=\tfrac12$, strictly increasing on $(\tfrac12,1)$, and consequently strictly decreasing on $(0,\tfrac12)$. This monotonicity holds for all $n$ when $\pi_0>\pi_1$, and for $\pi_0<\pi_1$ provided
\[
n>
\left(
\frac{
kc+\sqrt{k^2c^2+4mck}
}{
2mc
}
\right)^2,
\]
where
\[
m=2(\pi_1-\pi_0),\qquad
c=
\frac{
2z_{\alpha/2}(\pi_1-\pi_0)t\sqrt{D_0}
}{
D_1
}, \qquad
k=
\frac{
z_{\alpha/2}(d_1^2-d_0^2)t
}{
\sqrt{D_0}
},
\]
and
\[
d_j=2\pi_j-1,\qquad
D_j=1-d_j^2t^2,\quad j=0,1.
\]

\item For any prescribed privacy budget $\varepsilon_0>0$, the constrained optimization problem
\[
\max_{p\in(0,1)\setminus\left\{\tfrac12\right\}} \mathcal{P}(p)
\quad
\text{subject to}
\quad
\varepsilon(p)\le \varepsilon_0
\]
attains its optimum at a boundary point $p^\star$ satisfying
\[
\varepsilon(p^\star)=\varepsilon_0.
\]
\end{enumerate}
\end{theorem}

\begin{proof}
Referring to Table \ref{Table:RR-power}, the power function corresponding to Warner’s randomized response design can be written as
\begin{align}
\mathcal{P}(p)
&=
1+\Phi\!\left(A(p)-B(p)\right)
-\Phi\!\left(A(p)+B(p)\right),
\label{eq:warner_power}
\end{align}
where
\[
A(p)
=
\frac{
2\sqrt{n}(\pi_0-\pi_1)(2p-1)
}{
\sqrt{1-(2\pi_1-1)^2(2p-1)^2}
},
\]
and
\[
B(p)
=
\frac{
z_{\alpha/2}
\sqrt{1-(2\pi_0-1)^2(2p-1)^2}
}{
\sqrt{1-(2\pi_1-1)^2(2p-1)^2}
}.
\]

Let
\[
t=2p-1,
\qquad t\in(-1,1).
\]
Then $\mathcal{P}(\cdot)$ can be viewed as a function of $t$. Observe that $A(t)$ is odd and $B(t)$ is even. Using the identity
\[
\Phi(-x)=1-\Phi(x),
\]
we obtain
\begin{align*}
\mathcal{P}(-t)
&=
1+\Phi\!\left(-A(t)-B(t)\right)
-\Phi\!\left(-A(t)+B(t)\right)
\\
&=
1+\Phi\!\left(A(t)-B(t)\right)
-\Phi\!\left(A(t)+B(t)\right)
\\
&=
\mathcal{P}(t).
\end{align*}

Hence, $\mathcal{P}(t)$ is an even function, implying symmetry about $t=0$. Equivalently, $\mathcal{P}(p)$ is symmetric about $\frac{1}{2}$.

Differentiating \eqref{eq:warner_power} with respect to $t$ yields
\begin{align*}
\mathcal{P}'(t)
&=
\phi(A(t)-B(t))(A'(t)-B'(t))
-\phi(A(t)+B(t))(A'(t)+B'(t))
\notag
\\
&=
A'(t)\bigl[\phi(A(t)-B(t))-\phi(A(t)+B(t))\bigr]
-
B'(t)\bigl[\phi(A(t)-B(t))+\phi(A(t)+B(t))\bigr].
\label{eq:power_derivative}
\end{align*}

Under the stated conditions, it follows that (see Appendix \ref{Appendix_A} for details)
\[
\mathcal{P}'(t)>0,
\qquad t\in(0,1).
\]
Hence, $\mathcal{P}(t)$ is strictly increasing on $(0,1)$. Since $\mathcal{P}(t)$ is even, it follows immediately that $\mathcal{P}(t)$ is strictly decreasing on $(-1,0)$ under the same condition.

Since $t=2p-1$, the preceding monotonicity properties imply that $\mathcal{P}(p)$ is strictly increasing on $(\tfrac12,1)$ and strictly decreasing on $(0,\tfrac12)$.

For Warner’s randomized response design, the differential privacy budget is
\[
\varepsilon(p)
=
\max\left\{
\ln\frac{p}{1-p},
\ln\frac{1-p}{p}
\right\},
\qquad
p\in(0,1),\quad p\neq \frac12.
\]
Equivalently,
\[
\varepsilon(p)=
\begin{cases}
\displaystyle \ln\frac{p}{1-p},
& p\in(\tfrac12,1),
\\[2ex]
\displaystyle \ln\frac{1-p}{p},
& p\in(0,\tfrac12).
\end{cases}
\]
Hence, $\varepsilon(p)$ is strictly increasing on $(\tfrac12,1)$ and strictly decreasing on $(0,\tfrac12)$.

Now consider the constrained optimization problem
\[
\max_{p\in(0,1)\setminus\left\{\tfrac12\right\}} \mathcal{P}(p)
\quad
\text{subject to}
\quad
\varepsilon(p)\le \varepsilon_0.
\]

First, suppose $p\in(\tfrac12,1)$. Since $\varepsilon(p)$ is strictly increasing on this interval, the feasible set is of the form
\[
\{p\in(\tfrac12,1): \varepsilon(p)\le \varepsilon_0\}
=
(\tfrac12,p_0],
\]
where $p_0$ is uniquely determined by
\[
\varepsilon(p_0)=\varepsilon_0.
\]
Because $\mathcal{P}(p)$ is strictly increasing on $(\tfrac12,1)$, the maximum over the feasible region is attained at $p^\star=p_0$.

Next, suppose $p\in(0,\tfrac12)$. Since $\varepsilon(p)$ is strictly decreasing on this interval, the feasible set becomes
\[
\{p\in(0,\tfrac12): \varepsilon(p)\le \varepsilon_0\}
=
[p_0,\tfrac12),
\]
where $p_0$ satisfies
\[
\varepsilon(p_0)=\varepsilon_0.
\]
Since $\mathcal{P}(p)$ is strictly decreasing on $(0,\tfrac12)$, the maximum is again attained at $p^\star=p_0$.

Therefore, in both cases, the optimal solution occurs at the boundary of the feasible region and satisfies
\[
\varepsilon(p^\star)=\varepsilon_0.
\]
This completes the proof.
\end{proof}

\begin{corollary}
For Warner's randomized response design, let $\varepsilon_0 > 0$ denote a prescribed privacy budget. Then, the constraint
\[
\varepsilon(p) \le \varepsilon_0
\]
holds if and only if
\[
\frac{1}{1+e^{\varepsilon_0}}
\le p \le
\frac{e^{\varepsilon_0}}{1+e^{\varepsilon_0}}.
\]
\end{corollary}

\begin{corollary}
Let $\varepsilon_0 > 0$ be a prescribed privacy budget for Warner’s randomized response design. 
Under the constraint $\varepsilon(p) \le \varepsilon_0$, the optimal design parameter that 
maximizes the power $\mathcal{P}(p)$ is attained at the boundary of the feasible set and is given by
\[
p^\star \in \left\{ \frac{1}{1 + e^{\varepsilon_0}}, \ \frac{e^{\varepsilon_0}}{1 + e^{\varepsilon_0}} \right\}.
\]
Equivalently, the optimum is attained at the endpoints of the interval
\[
\left[ \frac{1}{1 + e^{\varepsilon_0}}, \ \frac{e^{\varepsilon_0}}{1 + e^{\varepsilon_0}} \right].
\]
\end{corollary}

\begin{corollary}[Power as a Function of the Privacy Budget]

Under the optimal choice of the design parameter, given by $p(\varepsilon) = \frac{e^{\varepsilon}}{1 + e^{\varepsilon}}$, the power of the test admits the following explicit representation as a function of the differential privacy budget $\varepsilon$.

\begin{align*}
\mathcal{P}(\varepsilon)
&=
1+\Phi\!\left(
\frac{2\sqrt{n}(\pi_0-\pi_1)\tanh(\frac{\varepsilon}{2})
-
z_{\alpha/2}\sqrt{1-(2\pi_0-1)^2\tanh^2(\frac{\varepsilon}{2})}
}
{\sqrt{1-(2\pi_1-1)^2\tanh^2(\frac{\varepsilon}{2})}}
\right) \\
&\quad
-\Phi\!\left(
\frac{2\sqrt{n}(\pi_0-\pi_1)\tanh(\frac{\varepsilon}{2})
+
z_{\alpha/2}\sqrt{1-(2\pi_0-1)^2\tanh^2(\frac{\varepsilon}{2})}
}
{\sqrt{1-(2\pi_1-1)^2\tanh^2(\frac{\varepsilon}{2})}}
\right).
\end{align*}
Moreover, $\mathcal{P}(\varepsilon)$ is strictly increasing in $\varepsilon$, reflecting the privacy-utility trade-off. The alternative choice $p=\frac{1}{1+e^{\varepsilon}}$ yields the same power.
\end{corollary}

\begin{remark}
The above theorem and its corollary are not specific to Warner’s randomized response design. The same conclusion holds for the Unrelated question design and the Two-Step randomized response design.

\end{remark}

\subsection{SAMPLE SIZE}

\begin{table}[h!]
    \renewcommand{\arraystretch}{2} 
    \centering
    \caption{Sample sizes for different randomized response designs.}
    \begin{adjustbox}{center}
    \resizebox{0.97\textwidth}{!}{%
    \begin{tabular}{|p{2cm}|p{12.2cm}|}
    \hline
    \multicolumn{1}{|c|}{\textbf{Design}} & \multicolumn{1}{c|}{\textbf{Sample Size ($n$)}} \\
    \hline
    \multicolumn{1}{|c|}{Warner} & \multicolumn{1}{|c|}{$\left(\frac{z_\beta \sqrt{\frac{1}{16(p - \frac{1}{2})^2} - (\pi_1 - \frac{1}{2})^2} + z_{\alpha/2} \sqrt{\frac{1}{16(p - \frac{1}{2})^2} - (\pi_0 - \frac{1}{2})^2}}{\pi_0 - \pi_1}\right)^2$}\\
    \hline
    \multicolumn{1}{|c|}{Unrelated Question} & \multicolumn{1}{|c|}{$\left(\frac{z_\beta \sqrt{\frac{[p \pi_1 + (1 - p) \pi_Y] - [p \pi_1 + (1 - p) \pi_Y]^2}{p^2}} + z_{\alpha/2} \sqrt{\frac{[p \pi_0 + (1 - p) \pi_Y] - [p \pi_0 + (1 - p) \pi_Y]^2}{p^2}}}{\pi_0 - \pi_1}\right)^2$}\\
    \hline
    \multicolumn{1}{|c|}{Forced Response} & \multicolumn{1}{|c|}{$\left(\frac{z_\beta \sqrt{\frac{[p_2 + (1 - p_1 - p_2) \pi_1] - [p_2 + (1 - p_1 - p_2) \pi_1]^2}{(1 - p_1 - p_2)^2}} + z_{\alpha/2} \sqrt{\frac{[p_2 + (1 - p_1 - p_2) \pi_0] - [p_2 + (1 - p_1 - p_2) \pi_0]^2}{(1 - p_1 - p_2)^2}}}{\pi_0 - \pi_1}\right)^2$}\\
    \hline
    \multicolumn{1}{|c|}{Kuk} & \multicolumn{1}{|c|}{$\left(\frac{z_\beta \sqrt{\frac{[p_1 \pi_1 + p_2 (1 - \pi_1)] - [p_1 \pi_1 + p_2 (1 - \pi_1)]^2}{(p_1 - p_2)^2}} + z_{\alpha/2} \sqrt{\frac{[p_1 \pi_0 + p_2 (1 - \pi_0)] - [p_1 \pi_0 + p_2 (1 - \pi_0)]^2}{(p_1 - p_2)^2}}}{\pi_0 - \pi_1}\right)^2$}\\
    \hline
    \multicolumn{1}{|c|}{Two-Step} & \multicolumn{1}{|c|}{$\left(\frac{z_\beta \sqrt{\frac{[p \pi_1 + p (1 - p)] - [p \pi_1 + p (1 - p)]^2}{p^2}} + z_{\alpha/2} \sqrt{\frac{[p \pi_0 + p (1 - p)] - [p \pi_0 + p (1 - p)]^2}{p^2}}}{\pi_0 - \pi_1}\right)^2$}\\
    \hline 
    \end{tabular}%
    }
    \end{adjustbox}
    \label{Table: RR-sample_size}
\end{table}

When the privacy budget is strictly specified as $\varepsilon(p) = c$, the corresponding value of the design parameter $p$ can be determined (uniquely for scalar $p$) as a function of the privacy budget. In this setting, the optimization problem reduces to examining whether the resulting $p$ yields sufficient statistical power for a given sample size $n$. Specifically, once $p$ is fixed by the privacy requirement, the focus shifts to finding the minimum sample size $n$ such that the power constraint $\Phi\big(d_1(n, p)\big) + 1 - \Phi\big(d_2(n, p)\big) \geq 1-\beta$ is satisfied. This approach allows practitioners to respect strict privacy guarantees while ensuring that the study remains adequately powered by appropriately adjusting the sample size. Investigators are generally most comfortable having an exact sample size formulation when designing a study, as this allows for precise planning and resource allocation. In practice, however, it is generally infeasible to solve the equation $\Phi\big(d_1(n, p)\big) + 1 - \Phi\big(d_2(n, p)\big) = 1 - \beta$ for $n$ analytically. Hence, we adopt an approximation based on the limiting behavior of the cumulative distribution function of the standard normal distribution. Recall that
\[
d_1 = \frac{\pi_0 - \pi_1 - z_{\alpha/2} \sqrt{V_{\pi_0}(T)}}{\sqrt{V_{\pi_1}(T)}}, \quad
d_2 = \frac{\pi_0 - \pi_1 + z_{\alpha/2} \sqrt{V_{\pi_0}(T)}}{\sqrt{V_{\pi_1}(T)}},
\]
where \( V_{\pi}(T) \) denotes the variance of the estimator \( T \) under the proportion \( \pi \), and \( z_{\frac{\alpha}{2}} \) is the upper \( \frac{\alpha}{2} \)-quantile of the standard normal distribution. When \( \pi_0 - \pi_1 < 0 \), for small value of $\alpha$, \( \Phi(d_1)\) is approximately 0, and the power equation simplifies to \( \Phi(d_2) = \beta \), which yields \( d_2 = z_{1 - \beta} \). Conversely, when \( \pi_0 - \pi_1 > 0 \), for small value of $\alpha$, \( \Phi(d_2)\) is approximately 1, and the equation reduces to \( \Phi(d_1) = 1 - \beta \), which gives \( d_1 = z_\beta \). In either case, the corresponding equation can be solved to obtain the required sample size \( n \), given the values of \( \pi_0 \), \( \pi_1 \), the design parameter \( p \), and the significance level \( \alpha \). Table \ref{Table: RR-sample_size} summarizes the approximate sample size expressions required to achieve power $1 - \beta$.


\section{SIMULATION STUDIES}\label{sec:simulation}
The simulation studies have three main objectives. First, we show that an optimal value of the design parameter $p$ may not always exist for a given power and privacy budget. This highlights the limitations of choosing $p$ arbitrarily and emphasizes the need for careful planning to balance power and privacy. Second, since collecting more samples is costly, the simulations compare different randomized response designs with respect to sample sizes. The goal is to identify which design requires the smallest sample size to achieve a specific power and privacy level. This helps researchers design studies that are both effective and efficient. The third objective is to examine how each randomized response design inflates the required minimum sample size to achieve stronger privacy protection relative to direct sampling.

\vspace{1cm}

\begin{figure*}[htbp]
\centering

\begin{subfigure}{0.47\textwidth}
    \centering
    \includegraphics[width=\linewidth]{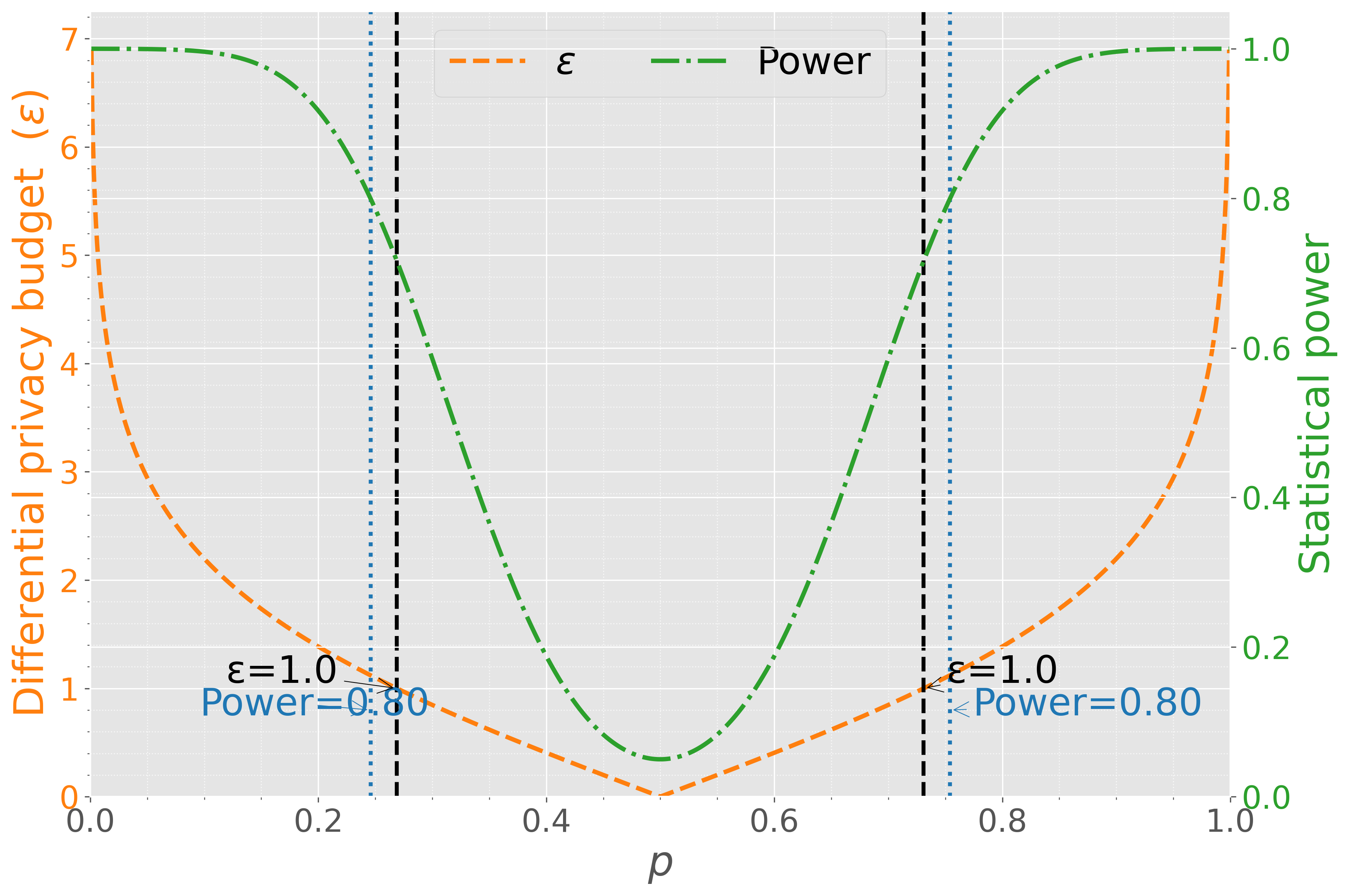}
    \caption{Warner's Design ($n = 700$)}
    \label{fig:warner_700}
\end{subfigure}\hfill
\begin{subfigure}{0.47\textwidth}
    \centering
    \includegraphics[width=\linewidth]{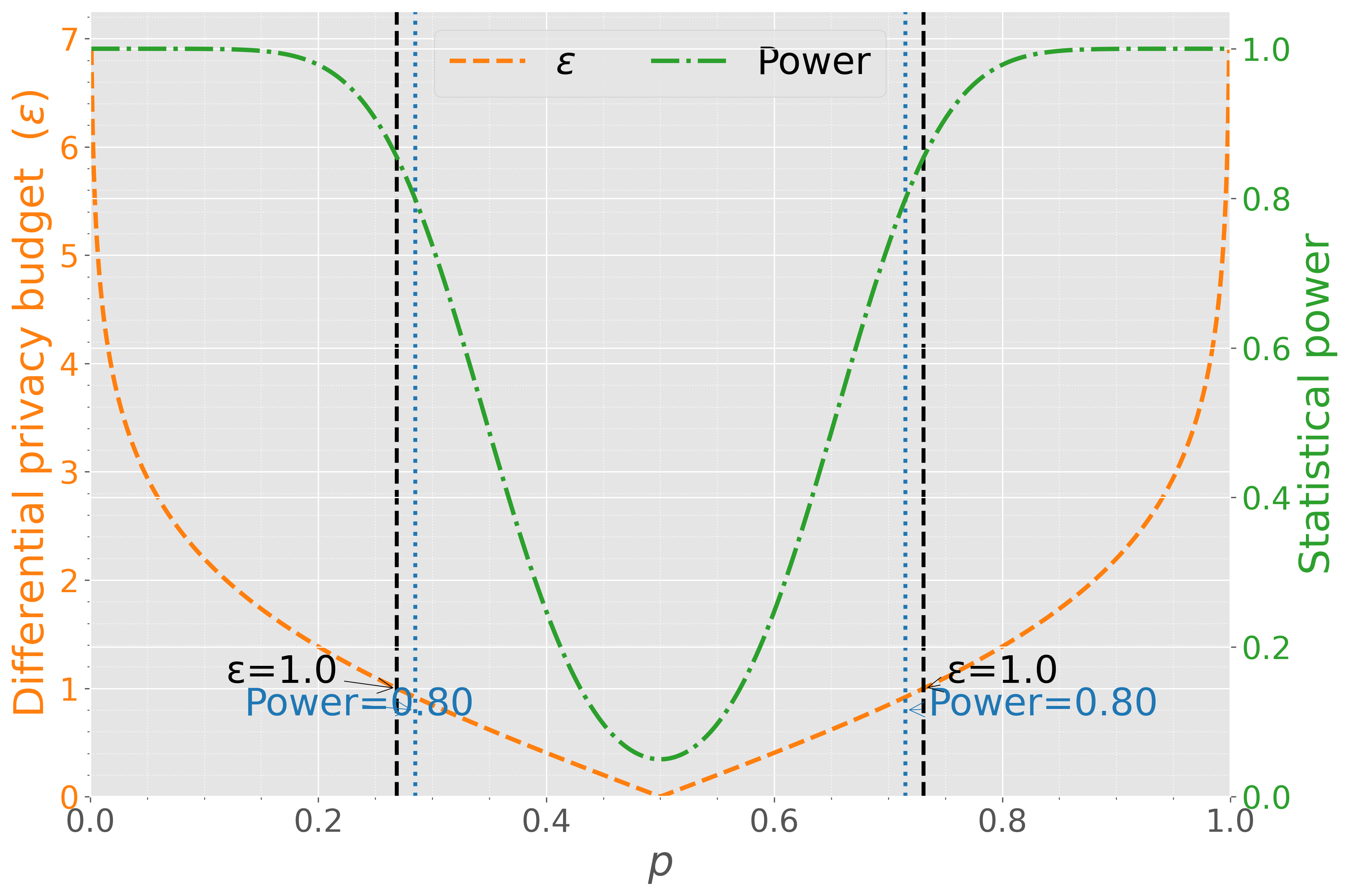}
    \caption{Warner's Design ($n = 1000$)}
    \label{fig:warner_1000}
\end{subfigure}





\end{figure*}



\begin{figure*}[htbp]\ContinuedFloat
\centering

\begin{subfigure}{0.42\textwidth}
    \centering
    \includegraphics[width=\linewidth]{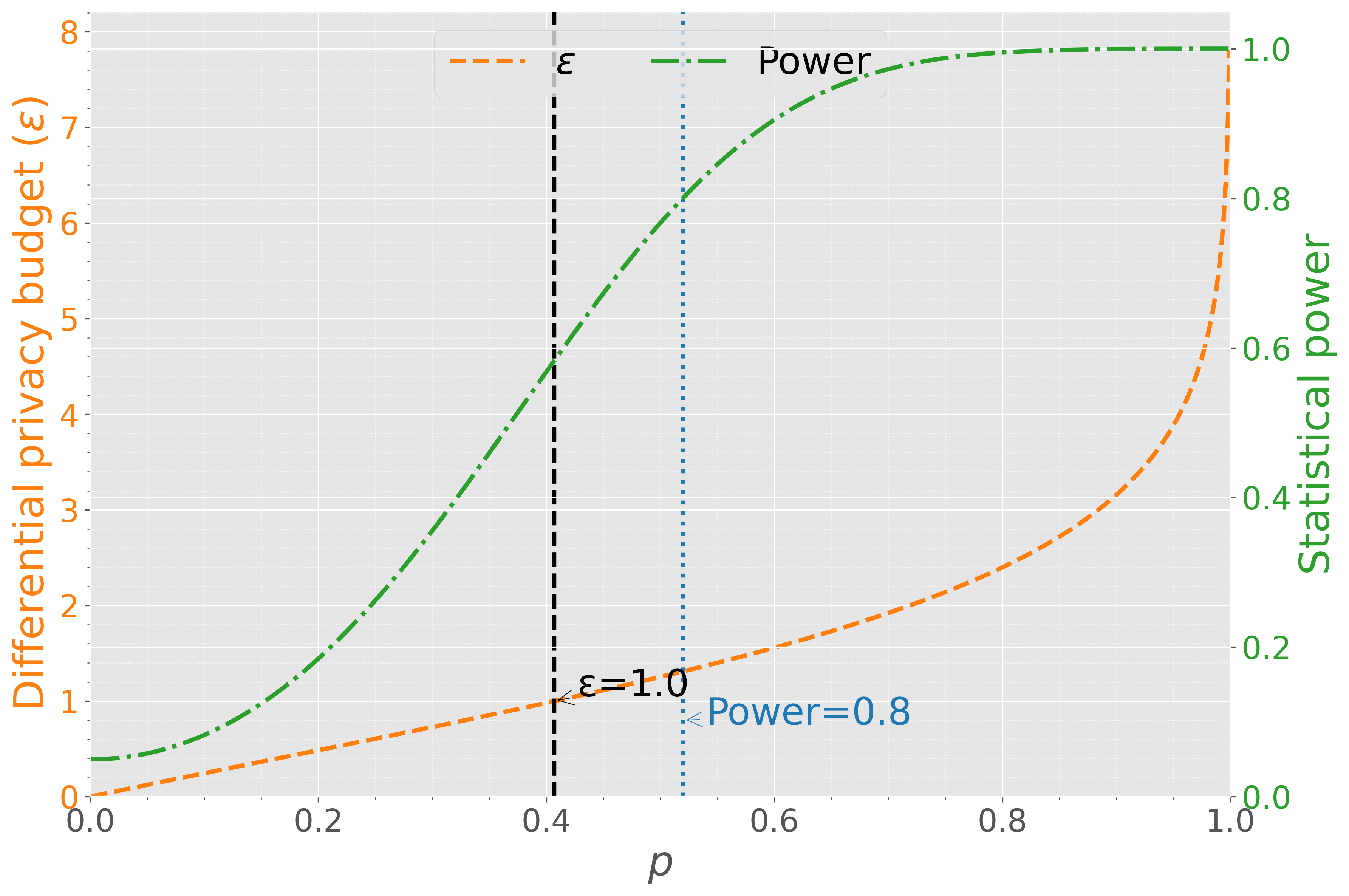}
    \caption{UQRR Design ($n = 700$)}
    \label{fig:uq_700}
\end{subfigure}\hfill
\begin{subfigure}{0.42\textwidth}
    \centering
    \includegraphics[width=\linewidth]{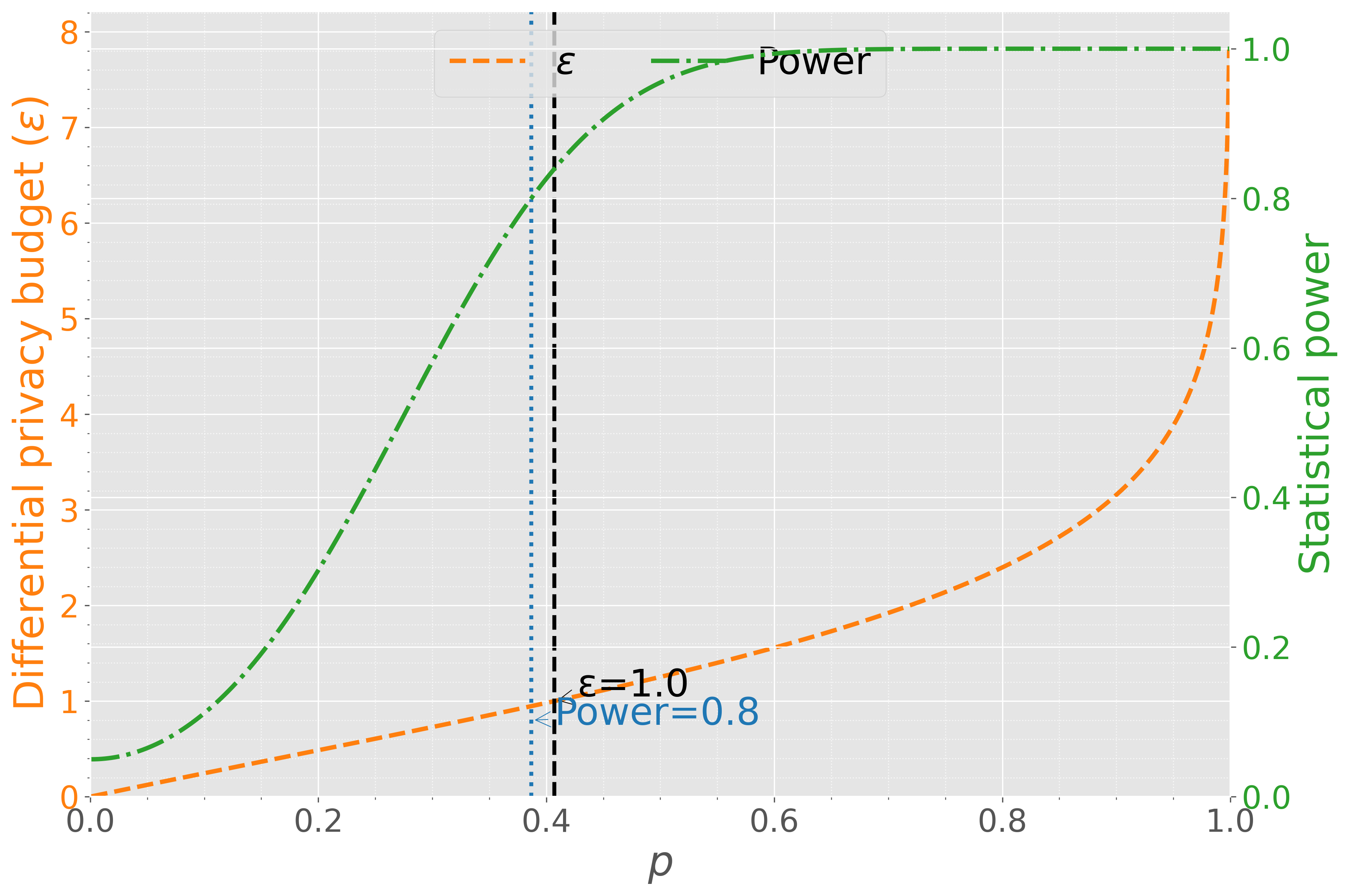}
    \caption{UQRR Design ($n = 1300$)}
    \label{fig:uq_1300}
\end{subfigure}

\begin{subfigure}{0.42\textwidth}
    \centering
    \includegraphics[width=\linewidth]{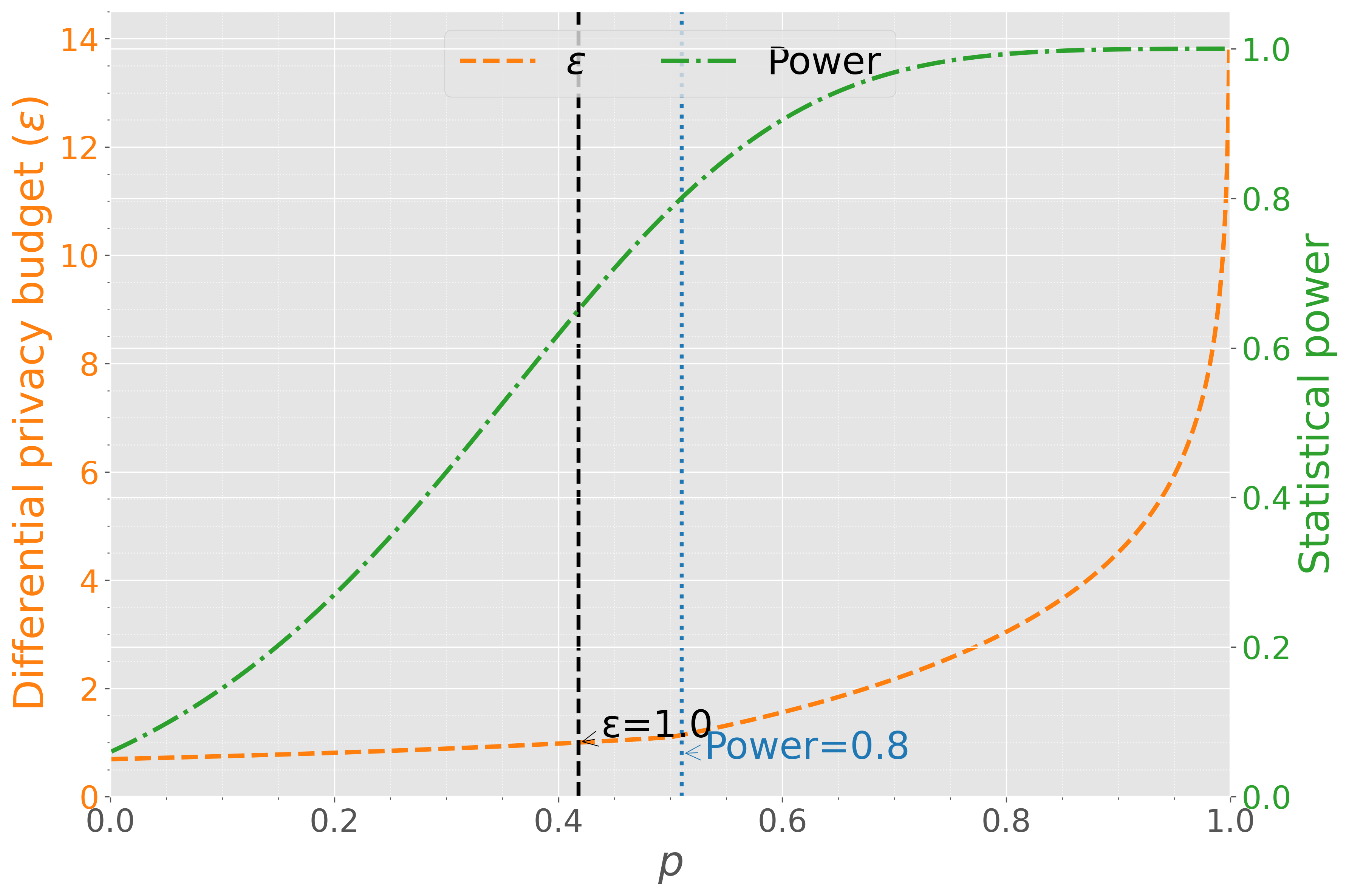}
    \caption{Two-Step Design ($n = 700$)}
    \label{fig:two_step_700}
\end{subfigure}\hfill
\begin{subfigure}{0.42\textwidth}
    \centering
    \includegraphics[width=\linewidth]{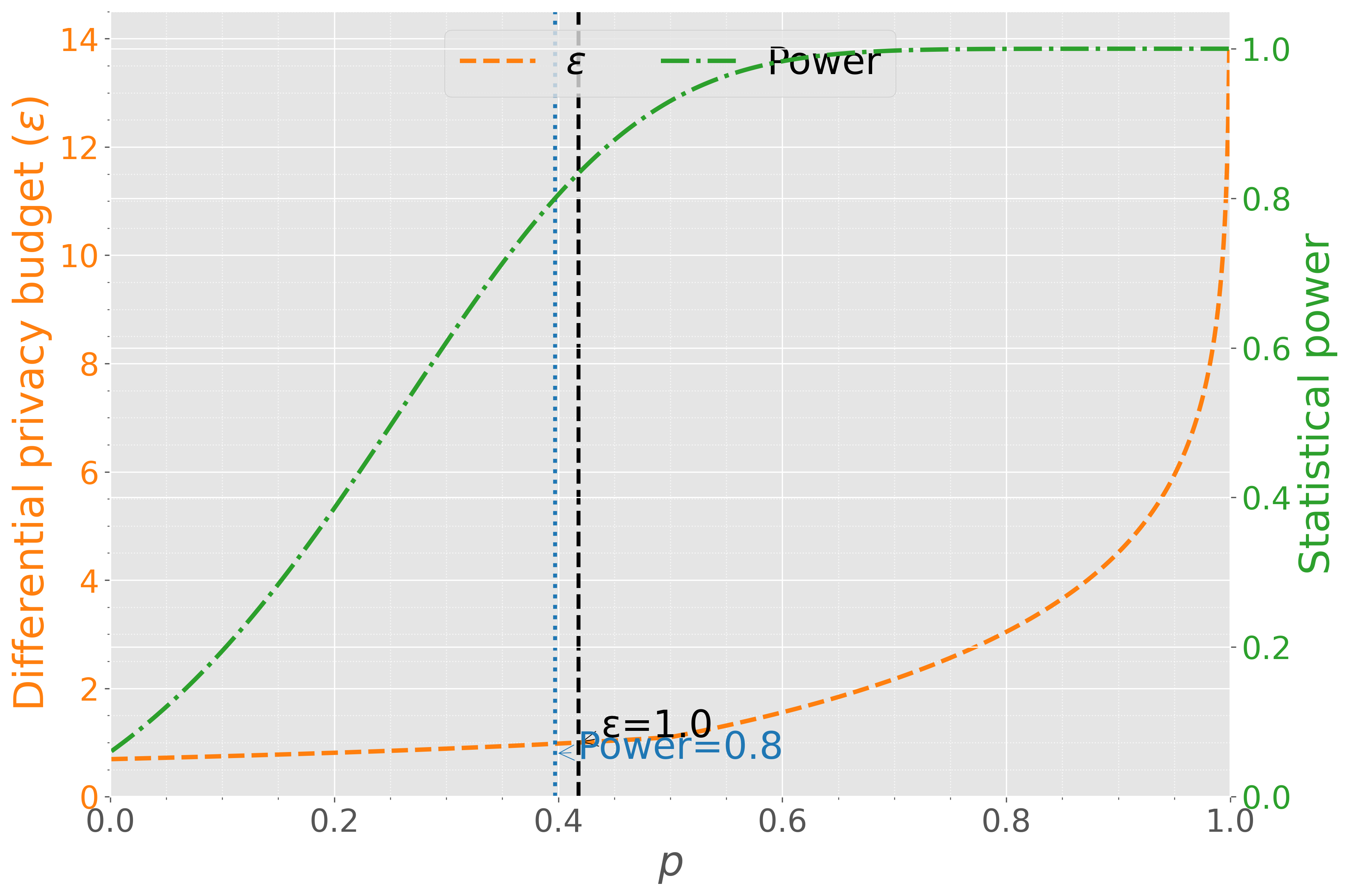}
    \caption{Two-Step Design ($n = 1100$)}
    \label{fig:two_step_1100}
\end{subfigure}


\begin{subfigure}{0.42\textwidth}
    \centering
    \includegraphics[width=\linewidth]{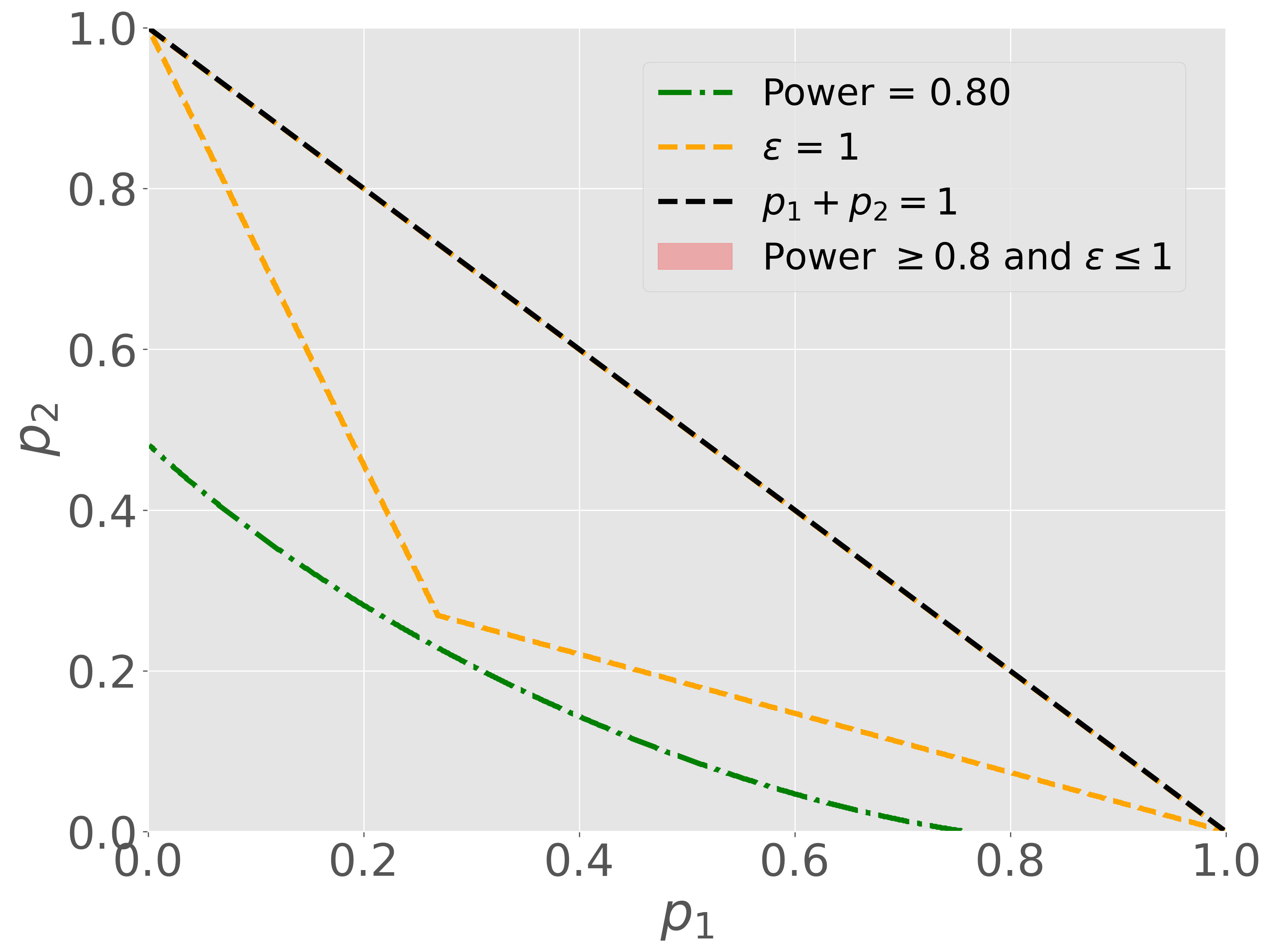}
    \caption{FRD ($n = 700$)}
    \label{fig:forced_700}
\end{subfigure}\hfill
\begin{subfigure}{0.42\textwidth}
    \centering
    \includegraphics[width=\linewidth]{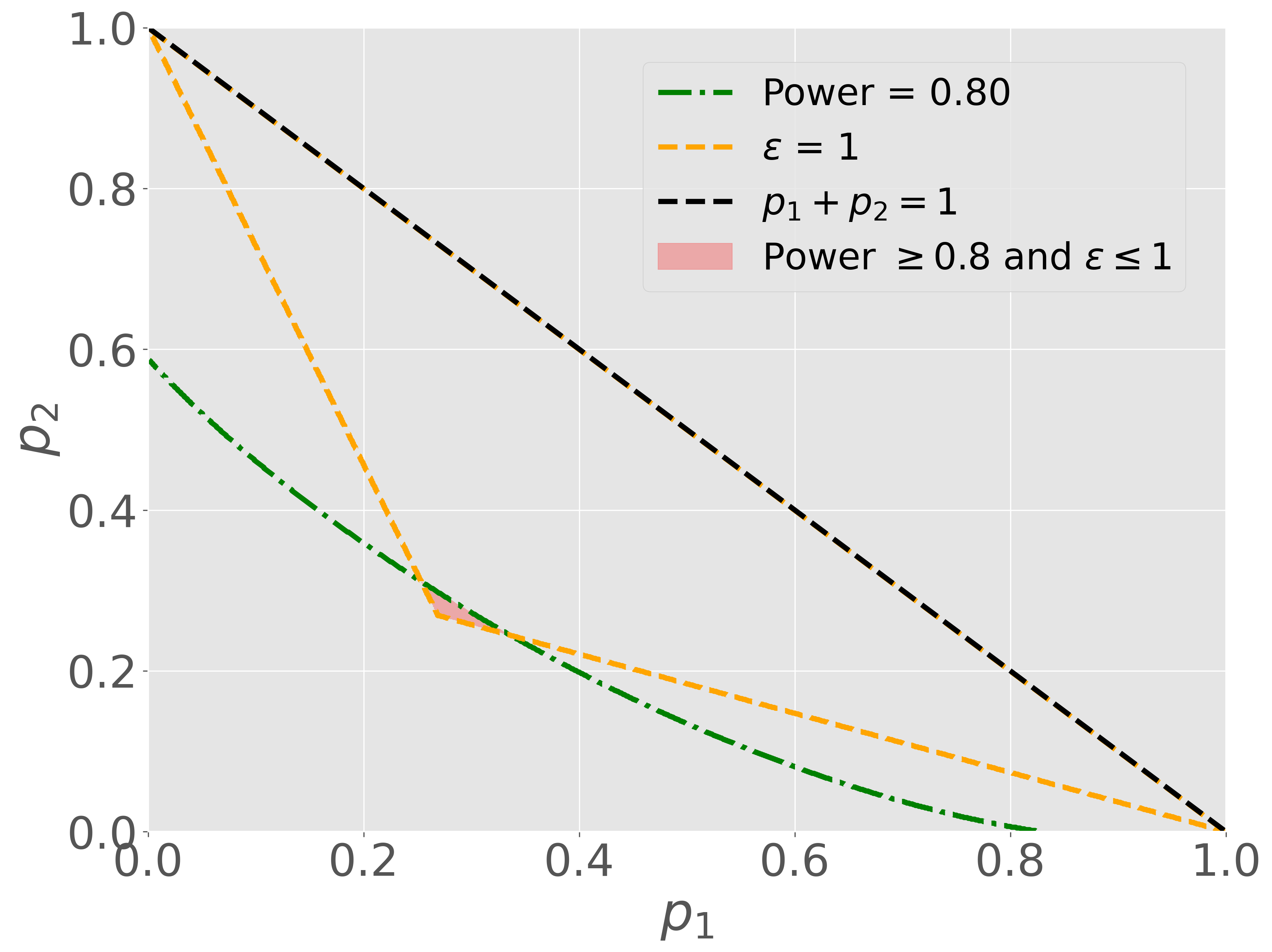}
    \caption{FRD ($n = 1000$)}
    \label{fig:forced_1000}
\end{subfigure}

\vspace{0.5cm}

\begin{subfigure}{0.42\textwidth}
    \centering
    \includegraphics[width=\linewidth]{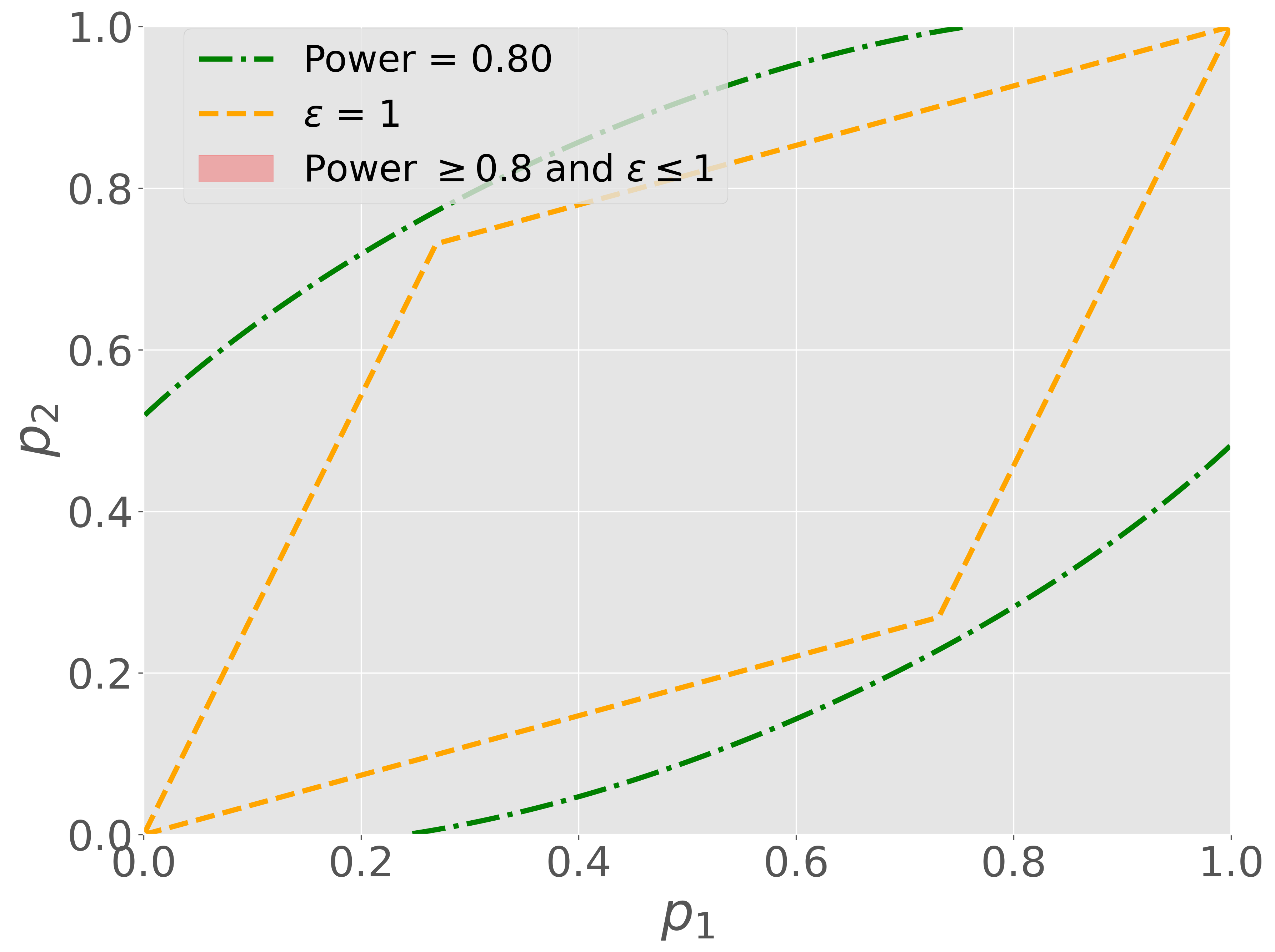}
    \caption{Kuk's Design ($n = 700$)}
    \label{fig:kuk_700}
\end{subfigure}\hfill
\begin{subfigure}{0.42\textwidth}
    \centering
    \includegraphics[width=\linewidth]{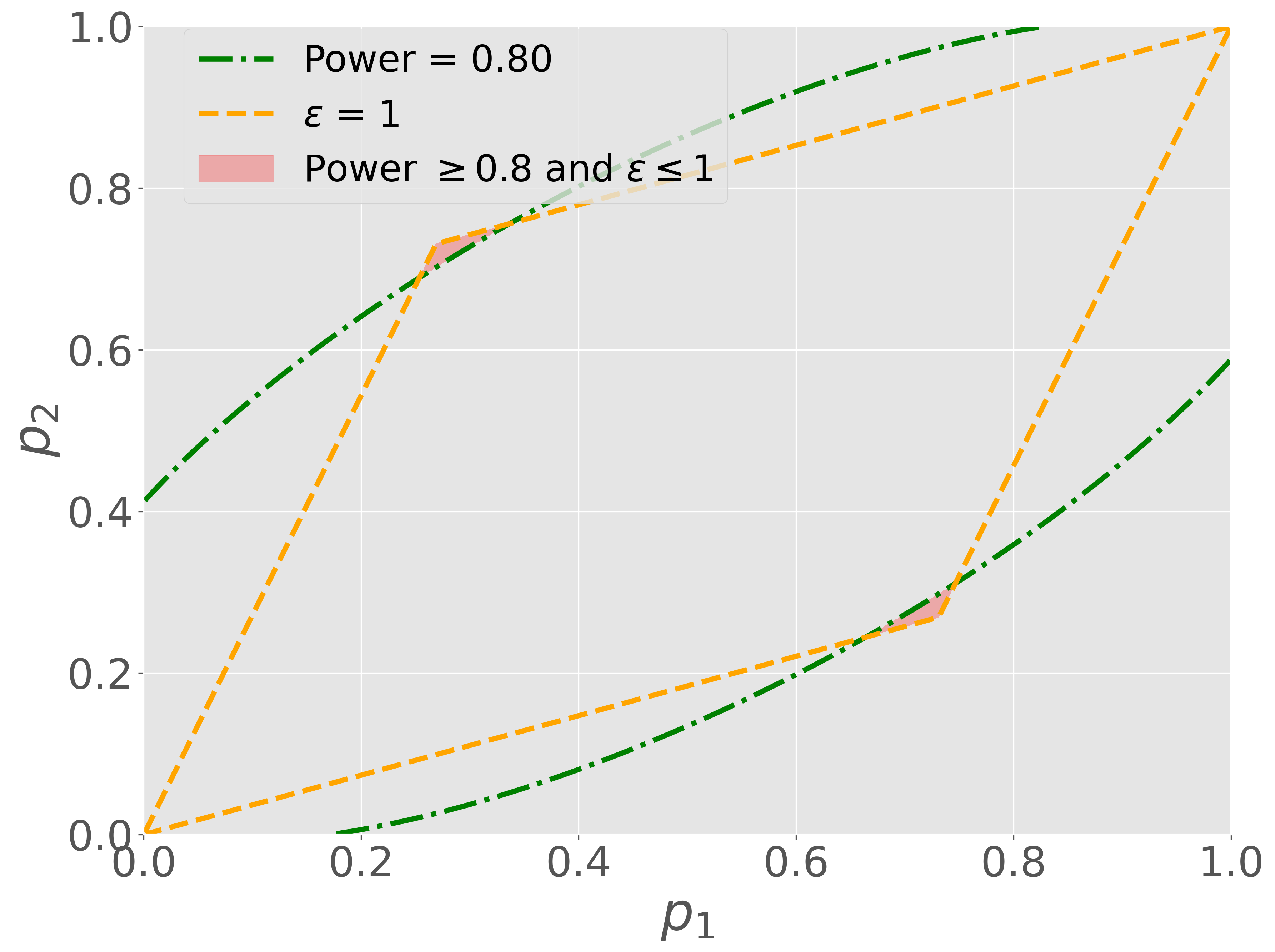}
    \caption{Kuk's Design ($n = 1000$)}
    \label{fig:kuk_1000}
\end{subfigure}

\caption{Illustration of optimal design parameters for different randomized response designs.}
\label{fig:rr_comparison}

\end{figure*}

\vspace{0.5cm}

Figure \ref{fig:rr_comparison} is organized into two columns and five rows, with each row representing a distinct RR design. The first column illustrates, for each RR design, a scenario in which an optimal value of the design parameter $p$ does not exist when the privacy budget is set at $\varepsilon = 1$ and the desired statistical power is $80\%$ with a fixed sample size. In the second column, when the sample size is increased, optimal value(s) exist for the design parameter $p$, with all other parameters held constant as in the corresponding figure in the first column. In Figure \ref{fig:rr_comparison}, we consider the parameter settings $\pi_0 = 0.2$, $\pi_1 = 0.3$, $\alpha = 0.05$, and $\pi_Y = 0.6$ (where $\pi_Y$ is used only in the unrelated question design). In the first column, the sample size is taken as $n=700$ for all five considered designs. In the first three rows, the left y-axis corresponds to $\varepsilon$, while the right y-axis represents the statistical power for Warner's design, the unrelated question design, and the two-step design, respectively.

In Warner's randomized response design (Figure \ref{fig:warner_700}), the privacy budget ($\varepsilon$, denoted by the orange dashed line) is minimized when $p$ is close to $0.5$ (note that here $p \neq 0.5$) and increases sharply as $p$ moves towards $0$ or $1$. In contrast, the power (denoted by the green dash-dot line) is lowest around $p = 0.5$ and increases toward the extremes. Blue dotted vertical lines mark the points $p=0.246$ and $p = 0.754$, where the power reaches approximately $0.80$, and black dashed vertical lines indicate $p = 0.269$ and $p = 0.731$, where $\varepsilon$ is close to $1$. From Figure \ref{fig:warner_700}, we observe that to ensure a privacy budget of $\varepsilon \leq 1$, the design parameter must lie within the interval $[0.269, 0.5)$ or $(0.5, 0.731]$. On the other hand, achieving a statistical power of at least $0.80$ requires $p \le 0.246$ or $p \ge 0.754$. There is no overlap of feasible values of $p$ corresponding to these two conditions. Therefore, for a sample size of $n=700$, there is no feasible value of the design parameter $p$ for which both $80\%$ power and a differential privacy budget of $\varepsilon \leq 1$ can be achieved. In Figure \ref{fig:warner_1000}, when the sample size is increased to $n = 1000$, optimal values of $p$ exist and belong to the sets $[0.269, 0.285]$ or $[0.715, 0.731]$ for Warner's design. Note that, with the given parameter setup, the minimum sample size is $n = 860$, at which an optimal value of $p$ exists. A larger sample size of $n = 1000$ is used to show that the optimal value of the parameter $p$ lies in the specified intervals and is not unique. Similarly, Figures \ref{fig:uq_700} and \ref{fig:two_step_700} show that there are no feasible values of the design parameter $p$ for the Unrelated Question and the Two-Step Randomized Response designs, respectively. However, when the sample sizes are set to $n = 1300$ and $n = 1100$ in Figures \ref{fig:uq_1300} and \ref{fig:two_step_1100}, optimal values of $p$ exist and belong to the sets $[0.387, 0.407]$ and $[0.397, 0.418]$, respectively.


\begin{figure*}[htbp]
\centering

\begin{subfigure}{0.47\textwidth}
    \centering
    \includegraphics[width=\linewidth]{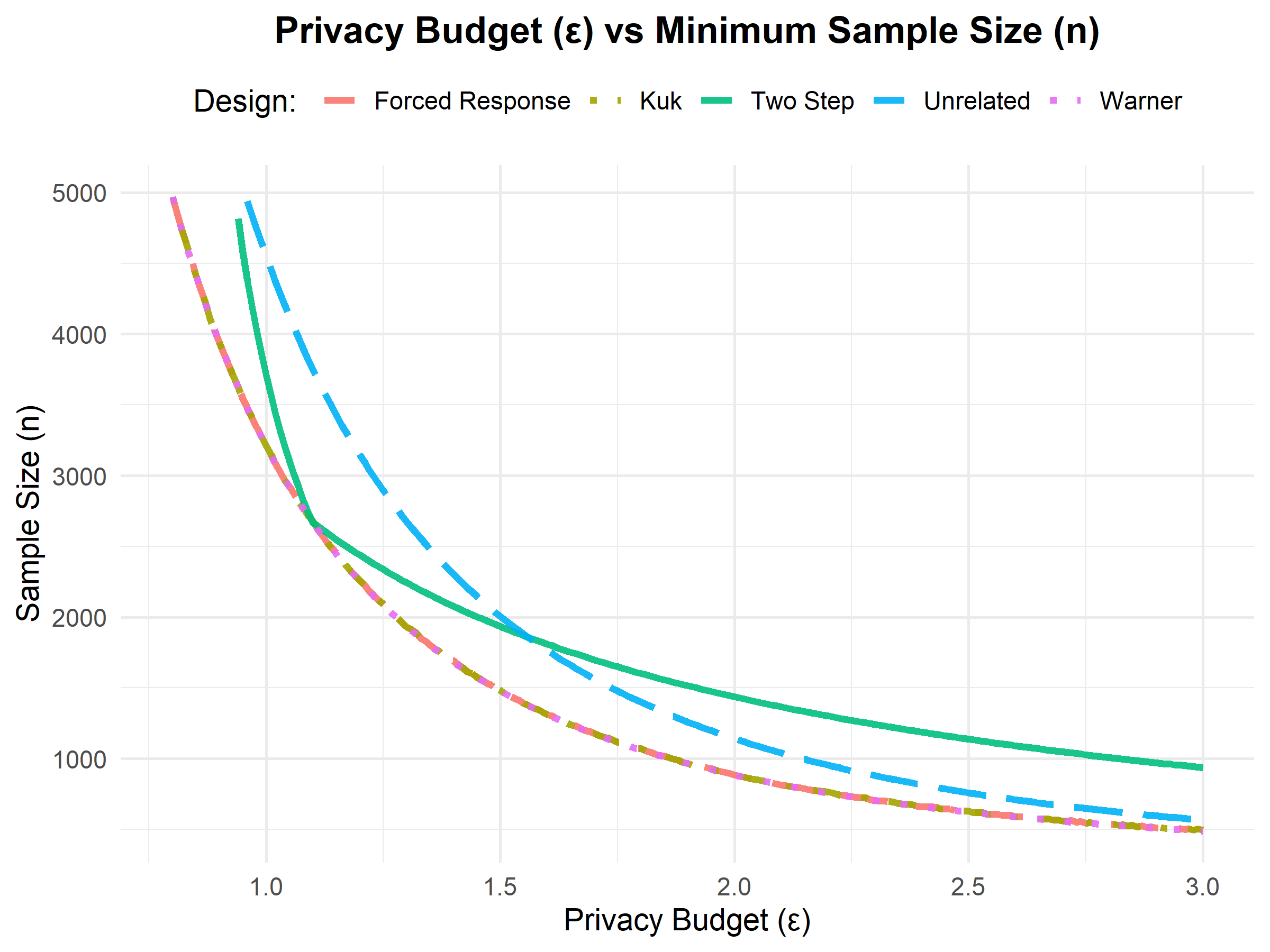}
    \caption{$\pi_0 = 0.1$, $\pi_1 = 0.15$, $\pi_Y = 0.6$}
    \label{fig:0.1vs0.15with0.6}
\end{subfigure}\hfill
\begin{subfigure}{0.47\textwidth}
    \centering
    \includegraphics[width=\linewidth]{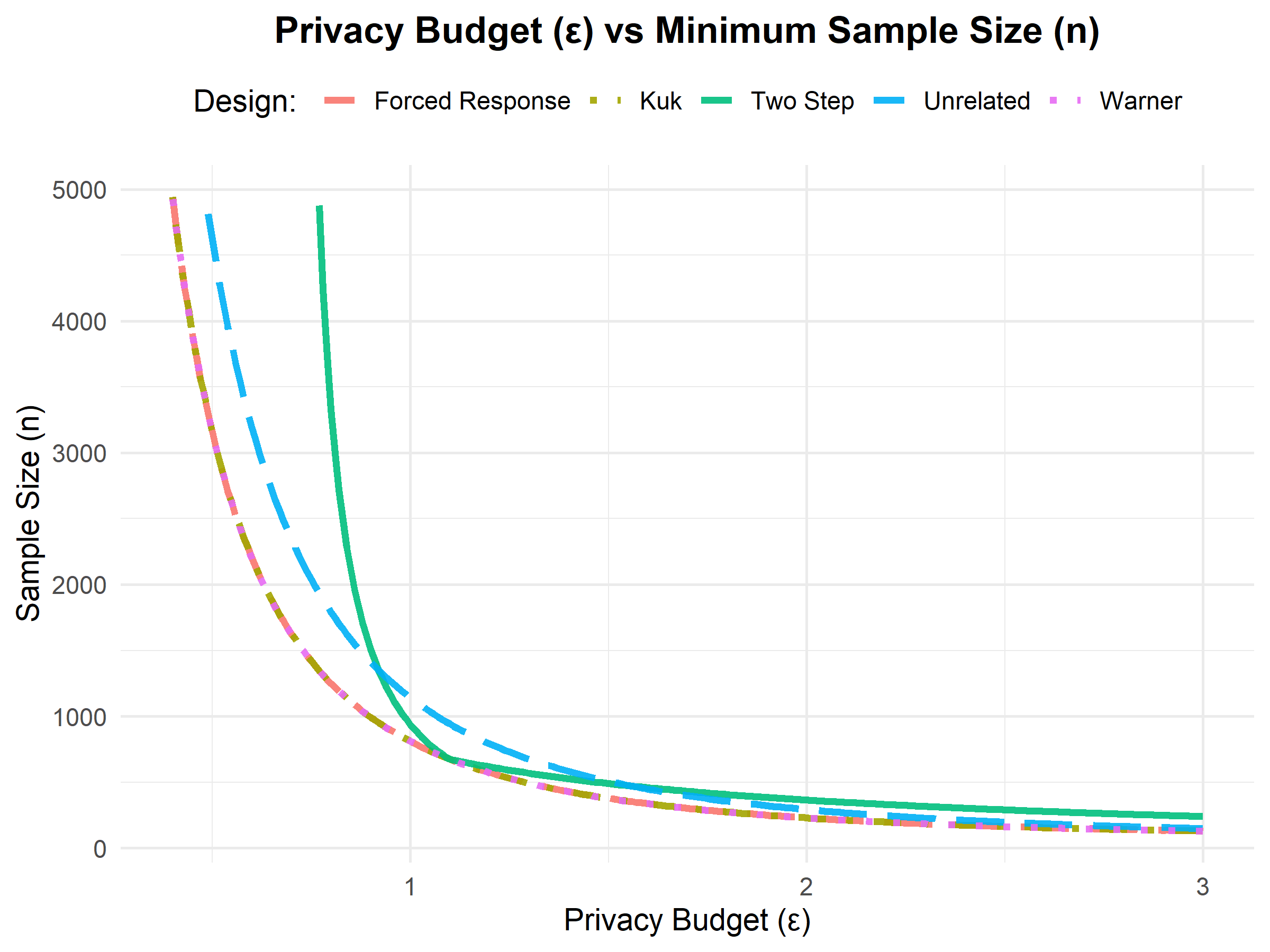}
    \caption{$\pi_0 = 0.1$, $\pi_1 = 0.2$, $\pi_Y = 0.6$}
    \label{fig:0.1vs0.2with0.6}
\end{subfigure}

\vspace{0.5cm}

\begin{subfigure}{0.47\textwidth}
    \centering
    \includegraphics[width=\linewidth]{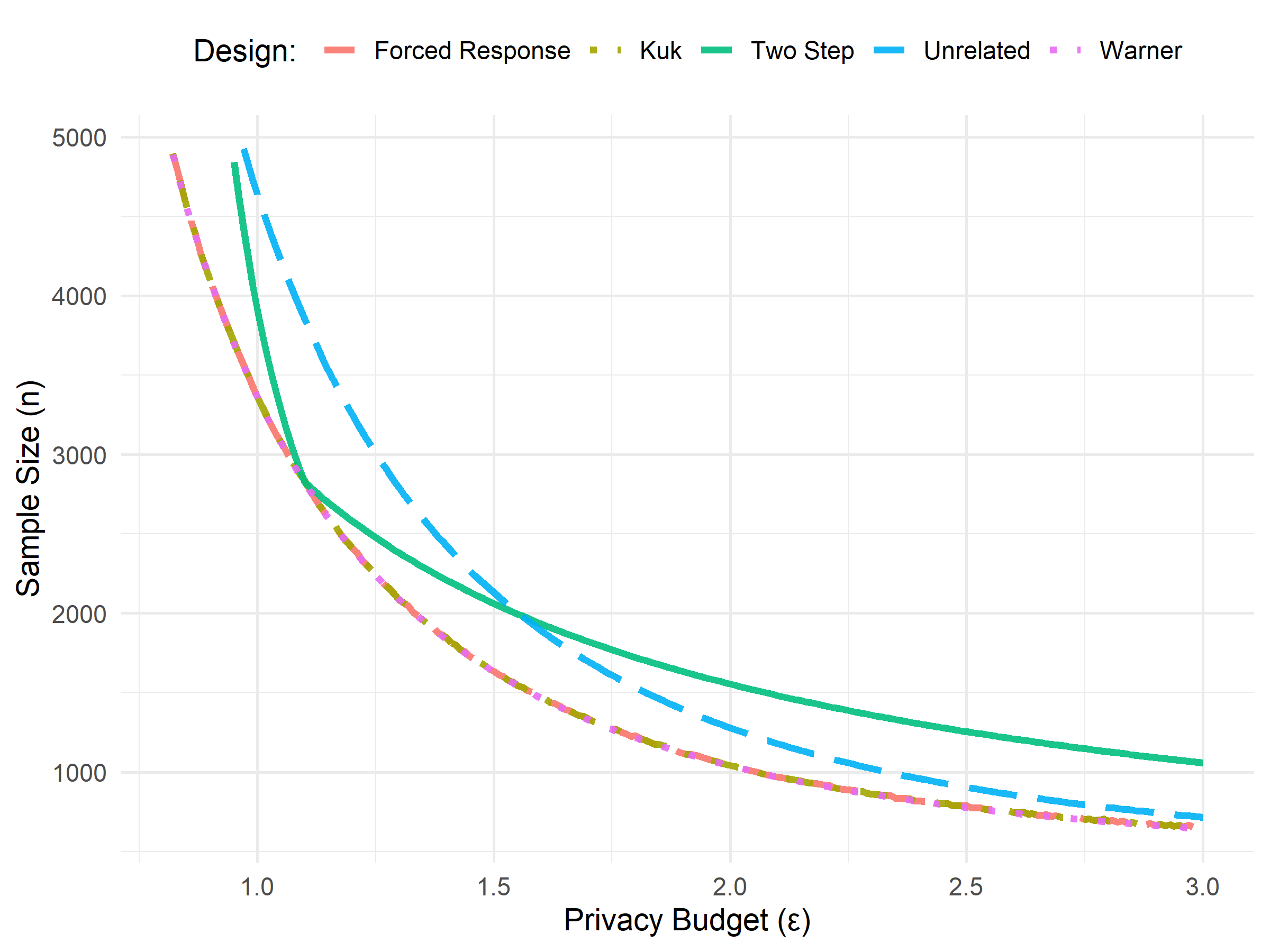}
    \caption{$\pi_0 = 0.2$, $\pi_1 = 0.15$, $\pi_Y = 0.6$}
    \label{fig:0.2vs0.15with0.6}
\end{subfigure}\hfill
\begin{subfigure}{0.47\textwidth}
    \centering
    \includegraphics[width=\linewidth]{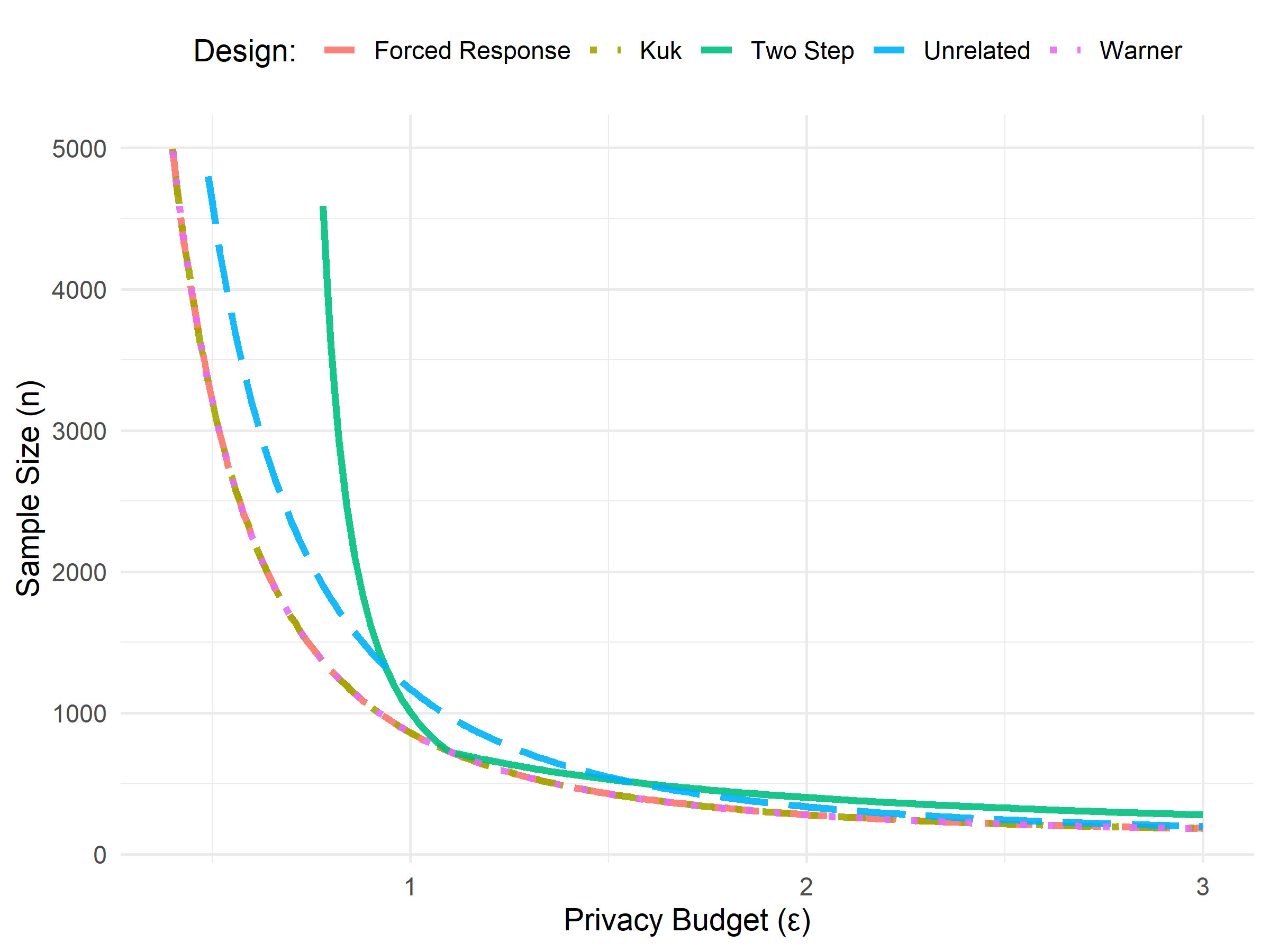}
    \caption{$\pi_0 = 0.2$, $\pi_1 = 0.3$, $\pi_Y = 0.6$}
    \label{fig:0.2vs0.3with0.6}
\end{subfigure}

\vspace{0.5cm}

\begin{subfigure}{0.47\textwidth}
    \centering
    \includegraphics[width=\linewidth]{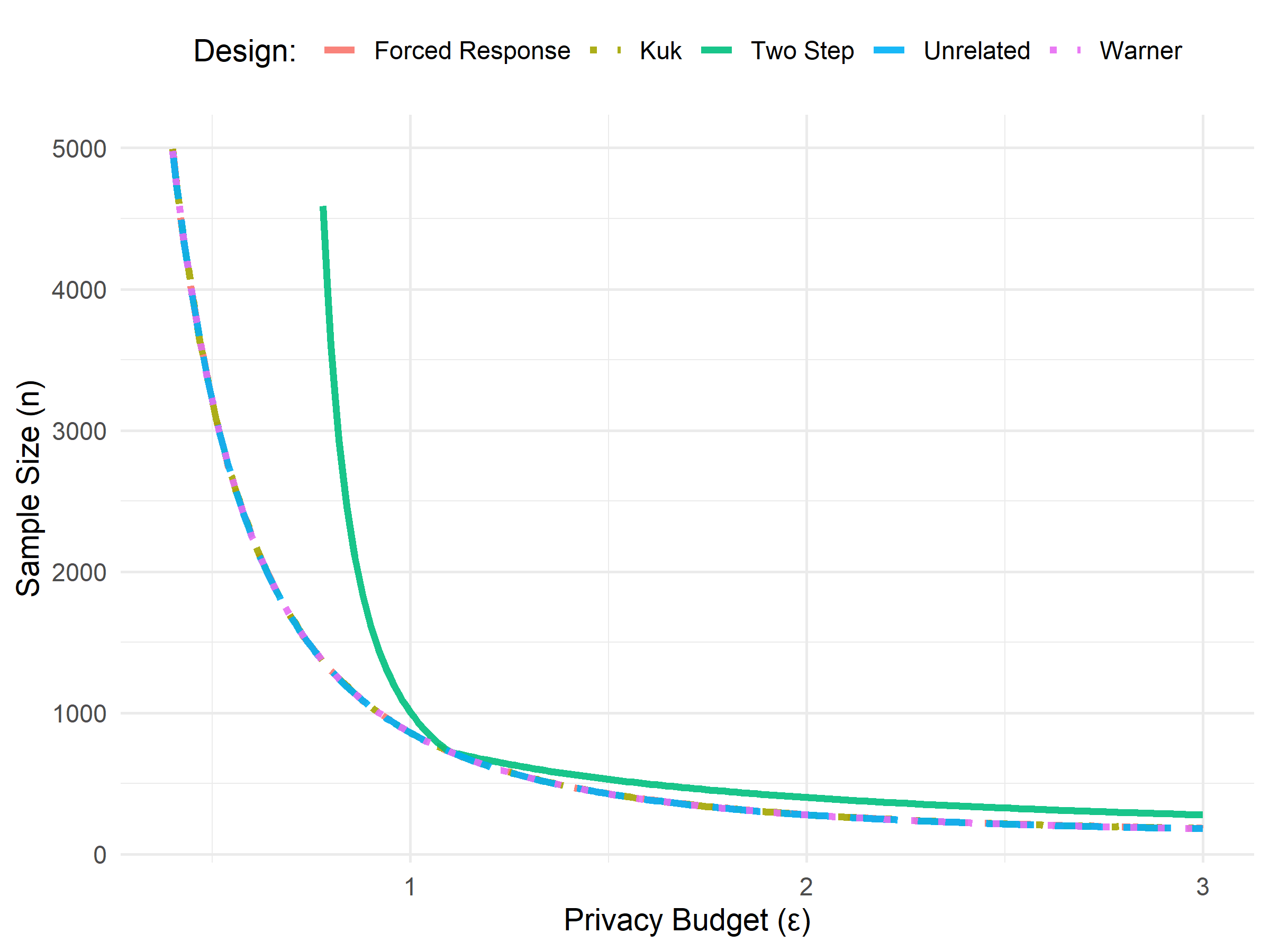}
    \caption{$\pi_0 = 0.2$, $\pi_1 = 0.3$, $\pi_Y = 0.5$}
    \label{fig:0.2vs0.3with0.5}
\end{subfigure}\hfill
\begin{subfigure}{0.47\textwidth}
    \centering
    \includegraphics[width=\linewidth]{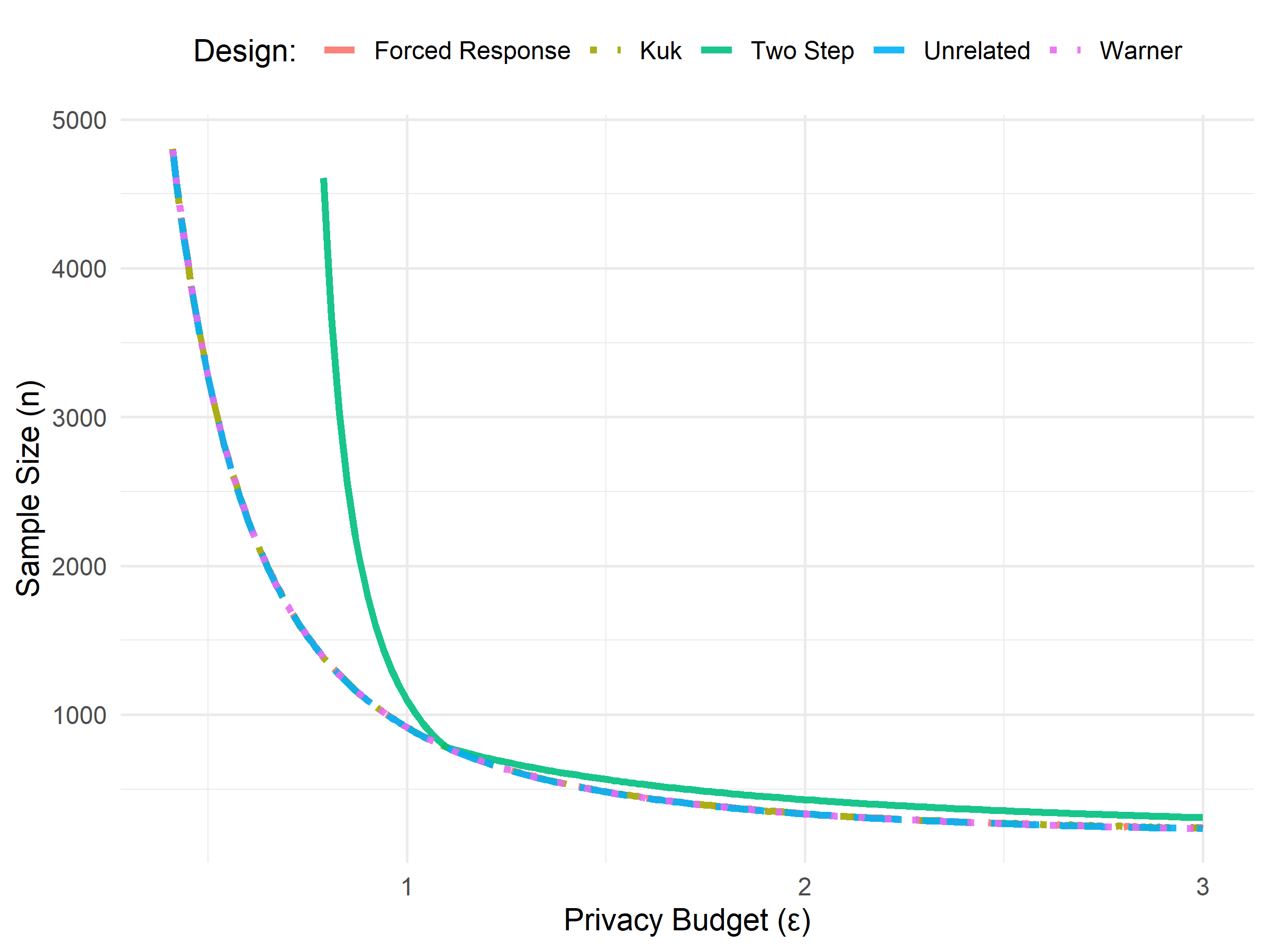}
    \caption{$\pi_0 = 0.4$, $\pi_1 = 0.5$, $\pi_Y = 0.5$}
    \label{fig:0.4vs0.5with0.5}
\end{subfigure}

\caption{Minimum sample size with respect to privacy budget ($\varepsilon$) for RR designs, holding power at 80\%}
\label{fig:privacy_vs_sample_size}

\end{figure*}

In the Forced Response and Kuk’s RR designs, the design parameter is a vector $p = (p_1, p_2)$. For the forced response design, along the orange dashed line (that is, where $\max \{ \ln \frac{1 - p_1}{p_2}, \ln \frac{1 - p_2}{p_1} \} = 1$ satisfying $p_1 + p_2 < 1$), the differential privacy budget is $\varepsilon = 1$. The privacy budget decreases as we move towards the line $p_1 + p_2 = 1$. On the green dash-dotted curve, the statistical power is $80\%$, and increases as we move closer to $(0, 0)$. From the Figure \ref{fig:forced_700}, it is evident that for $n = 700$, there is no feasible combination of $(p_1, p_2)$ for which the statistical power exceeds $80\%$ while the privacy budget satisfies $\varepsilon \leq 1$. When the sample size increases to $n = 1000$, in Figure \ref{fig:forced_1000}, we observe a feasible light-coral shaded region that represents the optimal values of $(p_1, p_2)$ where the power is at least $80\%$ and $\varepsilon \leq 1$. For example, $(p_1, p_2) = (0.335, 0.245)$, $(0.332, 0.246)$, and $(0.325, 0.249)$ lie within that light-coral shaded region. Similarly, there is no feasible region of optimal values of $(p_1, p_2)$ in Figure \ref{fig:kuk_700} for Kuk's design when the sample size is $ n =700$. However, there are two shared feasible light-coral shaded regions in Figure \ref{fig:kuk_1000} for Kuk's design when the sample size is increased to $n =1000$. In summary, this simulation study indicates that an optimal design-parameter value ($p$) may not exist when the sample size is limited.


\begin{table}[h!]
\renewcommand{\arraystretch}{1.8}
\centering
\caption{Comparison of required minimum sample sizes under direct sampling and optimal design (with parameters $\pi_0 = 0.2$, \(\pi_1 = 0.3\), \(\pi = 0.05\), \(\pi_Y = 0.6\)).}

\resizebox{0.97\textwidth}{!}{%
\begin{tabular}{|c|c|c|c|c|c|}
\hline
\multirow{2}{*}{Design} & 
\multicolumn{2}{c|}{\begin{tabular}[c]{@{}c@{}}Direct Sampling \\ 
(Power = 80\%)\end{tabular}} & 
\multicolumn{2}{c|}{\begin{tabular}[c]{@{}c@{}}Optimal Design \\ 
($\varepsilon = 1$, Power = 80\%)\end{tabular}} & 
\multirow{2}{*}{\shortstack{Inflation-Ratio\\($\frac{n_{RR}}{n_D}$)}} \\ \cline{2-5}
 & \begin{tabular}[c]{@{}c@{}}Design parameter \\ ($p$ or ($p_1,p_2$))\end{tabular} 
 & \begin{tabular}[c]{@{}c@{}}Sample Size \\ ($n_D$)\end{tabular}
 & \begin{tabular}[c]{@{}c@{}}Design parameter \\ ($p$ or ($p_1,p_2$))\end{tabular} 
 & \begin{tabular}[c]{@{}c@{}}Sample Size \\ ($n_{RR}$)\end{tabular} &  \\ \hline

Warner 
 & $1$ 
 & \multirow{5}{*}{137}
 & $0.269$ or $0.731$ 
 & $860$ & 6.28 \\ \cline{1-2} \cline{4-6}

Unrelated Question  
 & $1$
 &  
 & $0.407$ 
 & $1169$ & 8.53 \\ \cline{1-2} \cline{4-6}

Two-Step
 & $1$
 &  
 & $0.418$ 
 & $1006$ & 7.34 \\ \cline{1-2} \cline{4-6}

Forced Response  
 & $(0,0)$
 &  
 & $(0.269,0.269)$ 
 & $860$ & 6.28 \\ \cline{1-2} \cline{4-6}

Kuk 
 & $(1,0)$
 &  
 & $(0.731,0.269)$ 
 & $860$ & 6.28 \\ \hline

\end{tabular}%
}
\label{tab:sample_inflate}
\end{table}


For the second objective, we identify the most efficient randomized response (RR) design among the five considered, defined by the minimum sample size required to achieve the desired power and privacy budget ($\varepsilon$). In Figure \ref{fig:privacy_vs_sample_size}, power is fixed at 80\% for all panels. Figure \ref{fig:privacy_vs_sample_size} shows that the minimum-sample-size curves versus $\varepsilon$ for the Warner, Forced, and Kuk designs coincide, yielding a single curve that consistently lies below the corresponding curves for the two-step and unrelated designs in Figures \ref{fig:0.1vs0.15with0.6}, \ref{fig:0.1vs0.2with0.6}, \ref{fig:0.2vs0.15with0.6}, and \ref{fig:0.2vs0.3with0.6}. Therefore, with respect to minimum sample size, these three designs are equivalent; researchers may nonetheless select among them based on the intended application, considering implementation constraints and the design-specific limitations discussed in Section \ref{sec:RR-designs}. In Figure \ref{fig:0.1vs0.15with0.6}, more than 3,000 samples are required for $\varepsilon \le 1$ to attain 80\% power when $\pi_0 = 0.1$, $\pi_1 = 0.15$, and $\pi_Y = 0.6$ for the Warner, Forced, and Kuk designs. In Figure \ref{fig:0.1vs0.2with0.6}, with $\pi_1 = 0.2$ and other parameters as in Figure \ref{fig:0.1vs0.15with0.6}, the minimum sample size falls below 1,000 near $\varepsilon = 1$, as expected, as the difference between the $\pi_0$ and $\pi_1$ increased from 0.05 to 0.1. The curves for the two-step and unrelated designs cross near $\varepsilon = 1.5$ in Figure \ref{fig:0.1vs0.15with0.6} and near $\varepsilon = 1$ in Figure \ref{fig:0.1vs0.2with0.6}, indicating that, for larger $\varepsilon$, the unrelated design is preferable to the two-step design. In Figure \ref{fig:0.2vs0.3with0.5}, the minimum sample-size curves for all designs except the two-step design coincide and lie below that of the two-step design. In other words, when $\pi_Y = 0.5$, the unrelated design coincides with the other three designs, the Warner, Forced, and Kuk designs, with respect to minimum sample size.

Figure \ref{fig:privacy_vs_sample_size} shows that the required sample size rises sharply to achieve 80\% power when the privacy budget is near zero or below one. Randomized response designs increase the required minimum sample size relative to direct surveys without privacy protection. Focusing on a privacy budget of 1, we quantify this inflation by the ratio $\frac{n_{RR}}{n_D}$, where $n_{RR}$ is the minimum sample size under an RR design and $n_D$ is that under direct sampling. The second column of Table \ref{tab:sample_inflate} reports the design parameters that reduce each RR design to the direct survey design. The required minimum sample size for the direct survey design is 137. Warner’s, Forced Response, and Kuk’s designs yield the same $\frac{n_{RR}}{n_D}$, indicating comparable inflation at a privacy budget of 1. In contrast, the Two-step design has a ratio of 7.34 ($n_{RR} = 1006$), and the Unrelated Question design has a ratio of 8.53 ($n_{RR} = 1169$), making the latter the most inflationary for achieving the same power and privacy in the considered parameter setup.

\section{DATA ANALYSIS}\label{sec:data_analysis}
In the motivating example in Section \ref{sec: motivation_example}, we considered data from an AMT study that estimated the proportion of people who have ever provided misleading or incorrect information in their tax return. The forced response design used the following design parameters, $p_1 = 0.083$ and $p_2 = 0.167$. These values were chosen without considering the optimality criteria for the power or privacy budget. This led to a relatively high estimated privacy budget of 2.30, indicating the possibility of increased privacy leakage. In this section, we assess whether the study could have been designed optimally. Specifically, we examine whether the selection of design parameters using the proposed methodology can achieve the desired accuracy–privacy tradeoff with a smaller sample size. The randomized (masked) data collected under the forced-response design ($n = 1602$) cannot be used for this evaluation because the masking does not allow re-analysis under the same or alternative RR designs. Instead, we assess five RR designs with optimally chosen parameters using the direct-questioning data ($n = 809$) and compare the results with those from the original forced-response design. Here, for the power calculations, we take $\pi_0 = 0.1$ and $\pi_1 = 0.2$.

\begin{table}[ht]
\centering
\small
\renewcommand{\arraystretch}{1.25}
\caption{Analysis of Amazon Mechanical Turk (AMT) data ($n = 809$) under five randomized response designs using optimally chosen design parameters.}
\resizebox{0.97\textwidth}{!}{%
\begin{tabular}{llccccc}
\toprule
Scenario & Quantity 
& Warner & Unrelated question & Two-step & Forced response & Kuk \\ 
\midrule
\multirow{4}{*}{\shortstack{Fixed\\ privacy\\ budget\\ ($\varepsilon = 1$)}}
 & Optimal parameter & $0.269$ or $0.731$ & $0.407$ & $0.418$ & $(0.269,\,0.269)$ & $(0.731,\,0.269)$ \\
 & Estimate (SD) & $0.105$ (0.035) & $0.107$ (0.042) & $0.104$ (0.040) & $0.105$ (0.035) & $0.103$ (0.035) \\
 & Bias (MSE) & $0.001$ (0.001) & $0.003$ (0.002) & $0.000$ (0.002) & $0.001$ (0.001) & $-0.001$ (0.001) \\
 & Power & $0.800$ & $0.657$ & $0.742$ & $0.800$ & $0.800$ \\
\midrule
\multirow{4}{*}{\shortstack{Fixed\\ power\\ (80\%)}}
 & Optimal parameter & $0.269$ or $0.731$ & $0.477$ & $0.454$ & $(0.269,\,0.269)$ & $(0.269,\,0.731)$ \\
 & Estimate (SD) & $0.105$ (0.035) & $0.107$ (0.035) & $0.105$ (0.037) & $0.105$ (0.035) & $0.103$ (0.035) \\
 & Bias (MSE) & $0.001$ (0.001) & $0.003$ (0.001) & $0.001$ (0.001) & $0.001$ (0.001) & $-0.001$ (0.001) \\
 & Privacy budget ($\varepsilon$) & $1.000$ & $1.187$ & $1.040$ & $1.000$ & $1.000$ \\
\bottomrule
\end{tabular}%
}
\label{tab:data_analysis}
\end{table}

In Table \ref{tab:data_analysis}, we present two design-optimization scenarios for a fixed sample size ($n=809$). In the first row, we set the privacy budget to $\varepsilon=1$ and use (\ref{eq:single_para_p_optimization}) to identify the set of design-parameter values that satisfy the privacy constraint. From this set, we select the value of the design parameter that maximizes the power. In the second row, we target 80\% power and use (\ref{eq:single_para_p_optimization_fixed_power}) to identify the set of design-parameter values that meet the power requirement. From this set, we select the value of the design parameter that minimizes the privacy budget using (\ref{eq:single_para_p_optimization_min_privacy}).

In the first row, the forced response design with optimal parameter {\small $(p_1,p_2) = (0.269,0.269)$} achieves a test power of $0.80$ with the privacy budget fixed at $\varepsilon=1$. The sample size is 809, compared with 1602 in the original study, but the privacy budget is maintained at 1 rather than 2.30. This represents a substantial improvement in privacy protection. If the sample size is moderately increased (e.g., to about 900), the power exceeds 80\% while keeping $\varepsilon=1$. In the second row, the forced-response design with optimal parameters $(p_1,p_2)=(0.269,0.269)$ yields a minimized privacy budget of $1$ with the power fixed at 80\%. In both scenarios, the optimal value $(p_1,p_2)$ differ substantially from the original choices $(0.083,0.167)$. Moreover, in both cases, the estimated population proportion of the sensitive group, individuals who have ever provided misleading or incorrect information on their tax return, is close to the true value, with bias $\leq 0.003$ and mean squared error $\leq 0.002$. Overall, the proposed methodology designs forced-response studies with optimally selected parameters that achieve the desired power–privacy trade-off using smaller sample sizes than the original study. The gains are twofold, meeting power–privacy requirements and reducing sample size. Among the five RR designs reported in Table \ref{tab:data_analysis}, Warner’s design, Forced Response design, and Kuk’s design exhibit similar performance in both scenarios. In particular, each attains a power of approximately 80\% (rounded to three decimal places) in the first row and achieves a privacy budget of 1 (rounded to three decimal places) in the second row. 
All designs perform similarly in estimating the proportion of the sensitive group, as reflected by their comparable biases and MSEs. 


\section{SHINY APP}\label{sec:shiny_app}

\begin{figure}[h]
    \centering
       \includegraphics[width=\linewidth]{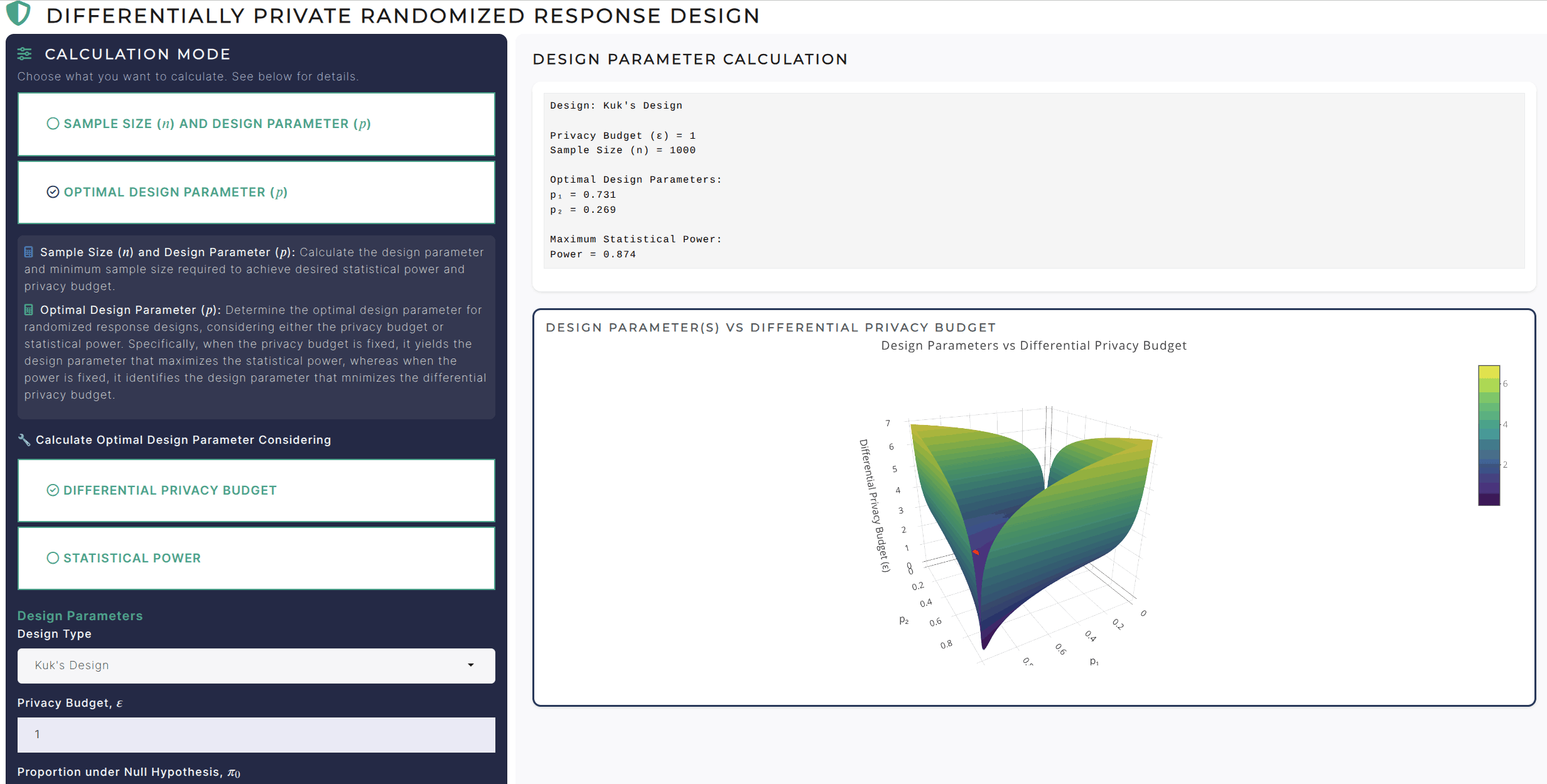}
   \caption{Screenshot of the Shiny application interface (\url{https://iitg.ac.in/pgapps/DP_RR/}).}

    \label{fig:shiny-app-screenshot}
\end{figure}

We developed a Shiny application, Differentially Private Randomized Response Design, that enables users to apply the proposed methods without writing code. The app is available at \url{https://iitg.ac.in/pgapps/DP_RR/} and provides an interactive platform for designing privacy-preserving randomized response studies. Figure \ref{fig:shiny-app-screenshot} shows a screenshot of the app. The interface is modern and user-friendly and supports two core functionalities: (1) calculating the required minimum sample size and corresponding optimal design parameters to achieve a specified statistical power under a given differential privacy budget $\varepsilon$; and (2) finding the value(s) of the design parameters either by fixing the privacy budget or by fixing the power for the five RR designs discussed in this work. In the second functionality, the app provides the design parameter that yields the maximum statistical power when the privacy budget is fixed. On the other hand, the app finds out the design parameter that ensures the minimum differential privacy budget when the power is fixed. These capabilities enable users to explore a wide range of scenarios without requiring knowledge of advanced statistical programming. The app automates complex computations and presents clear visualizations, including interactive 2D and 3D plots, thereby improving interpretability. It is particularly useful for studies involving sensitive or stigmatized topics, such as in public health, sociology, and behavioral research, where respondent privacy and truthful reporting are paramount. Because no single RR design is universally optimal, the app can be used to compare multiple designs under consistent settings to identify the design that best aligns with the study objectives.

\section{DISCUSSION}\label{sec:discussion}
Randomized response designs provide a rigorous and principled methodology for collecting data on sensitive topics while safeguarding respondents' privacy. Historically, RR design choices have tended to emphasize statistical power, with comparatively less explicit accounting for privacy guarantees. In contrast, work inspired by differential privacy in computer science prioritizes stringent privacy budgets but often overlooks power considerations that are central to inference. Our results demonstrate that privileging either objective in isolation can lead to RR designs that are suboptimal in practice, potentially compromising power or privacy to an unacceptable degree. We systematically characterize the trade-off between privacy budget and statistical power across RR designs and show how to navigate it using our proposed optimality criteria and accompanying algorithm. In particular, given a target privacy budget, our framework yields design parameters that attain a prespecified power, thereby enabling RR studies to meet both inferential and privacy goals.

In this work, we examine five well-known RR designs from both power and privacy perspectives. For each design, we provide an sample size formula and a ready-to-use Shiny app that computes sample size and optimal design parameters. Researchers can use the app to plan new RR studies efficiently. The app also allows side-by-side comparisons across designs, so users can select the RR design that achieves the same target power and privacy budget with the smallest required sample size.

In Figures \ref{fig:0.2vs0.3with0.5} and \ref{fig:0.4vs0.5with0.5}, we observe that for a fixed power level, the minimum required sample size remains the same for a fixed value of the differential privacy budget across the four designs, Warner, Unrelated Question, Forced Response, and Kuk’s designs. To understand the structural equivalence among these designs, we explore transformations of the design parameters $p$, or $(p_1, p_2)$, which produce identical sample size expressions across designs. Let $p_w$ denote the design parameter of Warner's RR design. For the Unrelated Question design, if $p$ is replaced by $(1 - 2p_w)$ for $p_w \in (0, \frac{1}{2})$ and by $(2p_w - 1)$ for $p_w \in (\frac{1}{2}, 1)$, and if $\pi_Y$ is fixed at $0.5$, the resulting sample size formula becomes identical to that of the Warner design. Similarly, in the Forced Response design, if both $p_1$ and $p_2$ are replaced by $p_w$ when $p_w \in (0, \frac{1}{2})$ and by $(1 - p_w)$ when $p_w \in (\frac{1}{2}, 1)$, the sample size expressions for the Forced Response and Warner designs coincide. Furthermore, in Kuk’s design, substituting $p_1$ with $p_w$ and $p_2$ with $(1 - p_w)$ leads to the same sample size as in Warner's design. These parameter transformations reveal underlying relationships among the designs, highlighting their equivalence in terms of required minimum sample size under comparable privacy and power constraints.

In the trade-off between power and privacy, we initially restricted the privacy budget and then optimized power with respect to design parameters (see equations (\ref{eq:privacy_budget-set}) and (\ref{eq:optimum_n_p})). We justified this decision by stating that the privacy budget is a function of the design parameter $p$ only, while power is a function of the parameters considered in the testing of the hypothesis, including the sample size ($n$) and the design parameter ($p$). However, there may be situations where privacy is the main concern; in such cases, it may be wise to restrict the power first and minimize the privacy budget to obtain the optimal values of the design parameter using equation (\ref{eq:single_para_p_optimization_fixed_power}).



\section*{Acknowledgement}
No potential competing interest was reported by the authors. This research was financially supported by the University Grants Commission (UGC), Government of India, through the award of the Research Fellowship under the National Eligibility Test (NET) (NTA Ref. No.: 221610111718) awarded to Bittu Karmakar. Palash Ghosh acknowledges the support from ANRF for this research under Grant No: ANRF/ARGM/2025/000054/MTR.




\section*{Appendix}
\label{Appendix_A}

The power function corresponding to Warner's randomized response design can be written as  
\begin{align*}
\mathcal{P}(p)
&=
1+\Phi\!\left(A(p)-B(p)\right)
-\Phi\!\left(A(p)+B(p)\right),
\end{align*}
where
\[
A(p)
=
\frac{
2\sqrt{n}(\pi_0-\pi_1)(2p-1)
}{
\sqrt{1-(2\pi_1-1)^2(2p-1)^2}
},
\]
and
\[
B(p)
=
\frac{
z_{\alpha/2}
\sqrt{1-(2\pi_0-1)^2(2p-1)^2}
}{
\sqrt{1-(2\pi_1-1)^2(2p-1)^2}
}.
\]

Let
\[
t=2p-1,
\qquad t\in(-1,1).
\]
Then the power function may be expressed as
\begin{align}
\mathcal{P}(t)
=
1+\Phi\!\left(A(t)-B(t)\right)
-\Phi\!\left(A(t)+B(t)\right),
\label{eq:power_t}
\end{align}
where
\[
A(t)
=
\frac{
2\sqrt{n}(\pi_0-\pi_1)t
}{
\sqrt{1-(2\pi_1-1)^2t^2}
},
\]
and
\[
B(t)
=
\frac{
z_{\alpha/2}
\sqrt{1-(2\pi_0-1)^2t^2}
}{
\sqrt{1-(2\pi_1-1)^2t^2}
}.
\]

Observe that $A(t)$ is an odd function of $t$, whereas $B(t)$ is an even function. Using the identity
\[
\Phi(-x)=1-\Phi(x),
\]
we obtain
\begin{align*}
\mathcal{P}(-t)
&=
1+\Phi\!\left(-A(t)-B(t)\right)
-\Phi\!\left(-A(t)+B(t)\right)
\\
&=
1+\bigl[1-\Phi(A(t)+B(t))\bigr]
-\bigl[1-\Phi(A(t)-B(t))\bigr]
\\
&=
1+\Phi(A(t)-B(t))
-\Phi(A(t)+B(t))
\\
&=
\mathcal{P}(t).
\end{align*}
Hence, $\mathcal{P}(t)$ is an even function, and therefore the power function is symmetric about $t=0$.

Differentiating \eqref{eq:power_t} with respect to $t$ yields
\begin{align}
\mathcal{P}'(t)
&=
\phi(A(t)-B(t))(A'(t)-B'(t))
-
\phi(A(t)+B(t))(A'(t)+B'(t))
\notag
\\
&=
A'(t)\bigl[\phi(A(t)-B(t))-\phi(A(t)+B(t))\bigr]
-
B'(t)\bigl[\phi(A(t)-B(t))+\phi(A(t)+B(t))\bigr].
\label{eq:power_derivative_general}
\end{align}

The derivatives of $A(t)$ and $B(t)$ are given by
\[
A'(t)
=
\frac{
2\sqrt{n}(\pi_0-\pi_1)
}{
\left[1-(2\pi_1-1)^2t^2\right]^{3/2}
},
\]
and
\[
B'(t)
=
\frac{
z_{\alpha/2}
\bigl[(2\pi_1-1)^2-(2\pi_0-1)^2\bigr]t
}{
\left[1-(2\pi_1-1)^2t^2\right]^{3/2}
\sqrt{1-(2\pi_0-1)^2t^2}
}.
\]

We now consider the monotonicity of $\mathcal{P}(t)$ on the interval $t\in(0,1)$.

\subsection*{Case 1: $\pi_0>\pi_1$}

For $t\in(0,1)$,
\[
A'(t)>0.
\]
Further, since $\pi_0>\pi_1$,
\[
(2\pi_0-1)^2>(2\pi_1-1)^2,
\]
which implies
\[
(2\pi_1-1)^2-(2\pi_0-1)^2<0.
\]
Hence,
\[
B'(t)<0.
\]

Moreover, for $t>0$,
\[
A(t)>0,
\qquad
B(t)>0,
\]

Since the standard normal density function $\phi(\cdot)$ is symmetric and strictly decreasing on $(0,\infty)$,
\[
\phi(A(t)-B(t))-\phi(A(t)+B(t))>0.
\]
Also,
\[
\phi(A(t)-B(t))+\phi(A(t)+B(t))>0.
\]

Substituting these sign relations into \eqref{eq:power_derivative_general}, we obtain
\[
\mathcal{P}'(t)>0,
\qquad t\in(0,1).
\]
Thus, the power function is strictly increasing on $(0,1)$ whenever $\pi_0>\pi_1$.

\subsection*{Case 2: $\pi_0<\pi_1$}

When $\pi_0<\pi_1$, we have $A(t)<0$ for $t\in(0,1)$. Define
\[
A_1(t)=-A(t).
\]
Then $A_1(t)>0$, and the power function can be rewritten as
\begin{align*}
\mathcal{P}(t)
&=
1+\Phi(-A_1(t)-B(t))-\Phi(-A_1(t)+B(t))
\\
&=
1+\bigl[1-\Phi(A_1(t)+B(t))\bigr]
-\bigl[1-\Phi(A_1(t)-B(t))\bigr]
\\
&=
1+\Phi(A_1(t)-B(t))-\Phi(A_1(t)+B(t)).
\end{align*}

Differentiating with respect to $t$ gives
\begin{align}
\mathcal{P}'(t)
&=
\phi(A_1(t)-B(t))(A_1'(t)-B'(t))
-
\phi(A_1(t)+B(t))(A_1'(t)+B'(t))
\notag
\\
&=
A_1'(t)
\bigl[\phi(A_1(t)-B(t))-\phi(A_1(t)+B(t))\bigr]
-
B'(t)
\bigl[\phi(A_1(t)-B(t))+\phi(A_1(t)+B(t))\bigr].
\label{eq:power_derivative_case2}
\end{align}

Using
\[
\phi(x)
=
\frac{1}{\sqrt{2\pi}}e^{-x^2/2},
\]
equation \eqref{eq:power_derivative_case2} simplifies to
\begin{align}
\mathcal{P}'(t)
=
\frac{
2e^{-(A_1^2(t)+B^2(t))/2}
}{
\sqrt{2\pi}
}
\left[
A_1'(t)\sinh(A_1(t)B(t))
-
B'(t)\cosh(A_1(t)B(t))
\right].
\label{eq:power_derivative_hyperbolic}
\end{align}

Since the exponential factor and $\cosh(A_1(t)B(t))$ are positive, it follows that
\[
\mathcal{P}'(t)>0
\]
whenever
\[
A_1'(t)\tanh(A_1(t)B(t))>B'(t).
\]

Let
\[
d_0=2\pi_0-1,
\qquad
d_1=2\pi_1-1,
\]
and define
\[
D_0=1-d_0^2t^2,
\qquad
D_1=1-d_1^2t^2.
\]
Then
\[
A_1'(t)
=
\frac{
2\sqrt{n}(\pi_1-\pi_0)
}{
D_1^{3/2}
},
\]
and
\[
B'(t)
=
\frac{
z_{\alpha/2}(d_1^2-d_0^2)t
}{
D_1^{3/2}\sqrt{D_0}
}.
\]

Consequently,
\[
\mathcal{P}'(t)>0
\]
provided that
\begin{align}
2(\pi_1-\pi_0)\sqrt{n}\,
\tanh\!\left(
\frac{
2z_{\alpha/2}\sqrt{n}(\pi_1-\pi_0)t\sqrt{D_0}
}{
D_1
}
\right)
>
\frac{
z_{\alpha/2}(d_1^2-d_0^2)t
}{
\sqrt{D_0}
}.
\label{eq:sufficient_condition}
\end{align}

Now define
\[
m=2(\pi_1-\pi_0),
\]
\[
c=
\frac{
2z_{\alpha/2}(\pi_1-\pi_0)t\sqrt{D_0}
}{
D_1
},
\]
and
\[
k=
\frac{
z_{\alpha/2}(d_1^2-d_0^2)t
}{
\sqrt{D_0}
}.
\]
Then condition \eqref{eq:sufficient_condition} becomes
\[
m\sqrt{n}\,\tanh(c\sqrt{n})>k.
\]

Using the inequality
\[
\tanh(x)\ge \frac{x}{1+x},
\qquad x\ge0,
\]
we obtain the sufficient condition
\[
m\sqrt{n}
\left(
\frac{c\sqrt{n}}{1+c\sqrt{n}}
\right)
>k.
\]
After simplification,
\[
mc\,n-kc\sqrt{n}-k>0.
\]

Let $y=\sqrt{n}$. Then
\[
mc\,y^2-kc\,y-k>0.
\]
Solving the corresponding quadratic equation yields
\[
y>
\frac{
kc+\sqrt{k^2c^2+4mck}
}{
2mc
}.
\]
Therefore, a sufficient condition for the monotonicity of the power function is
\[
n>
\left(
\frac{
kc+\sqrt{k^2c^2+4mck}
}{
2mc
}
\right)^2.
\]

Hence, for $t\in(0,1)$ and $\pi_0<\pi_1$, the power function $\mathcal{P}(t)$ is increasing whenever
\[
n>
\left(
\frac{
kc+\sqrt{k^2c^2+4mck}
}{
2mc
}
\right)^2.
\]

Because $\mathcal{P}(t)$ is an even function, we have
\[
\mathcal{P}(-t)=\mathcal{P}(t),
\qquad t\in(-1,1).
\]
Hence, wherever $\mathcal{P}$ is differentiable,
\[
\mathcal{P}'(-t)=-\mathcal{P}'(t).
\]
Therefore, the behavior of $\mathcal{P}(t)$ on $(-1,0)$ is the mirror image of its behavior on $(0,1)$.

In particular, when $\pi_0>\pi_1$, it was shown previously that
\[
\mathcal{P}'(t)>0,
\qquad t\in(0,1).
\]
Therefore,
\[
\mathcal{P}'(t)<0,
\qquad t\in(-1,0),
\]
which implies that $\mathcal{P}(t)$ is strictly decreasing on $(-1,0)$.

Similarly, when $\pi_0<\pi_1$, under the sufficient condition
\[
n>
\left(
\frac{
kc+\sqrt{k^2c^2+4mck}
}{
2mc
}
\right)^2,
\]
it was established that $\mathcal{P}(t)$ is strictly increasing on $(0,1)$. By symmetry, it follows that
\[
\mathcal{P}'(t)<0,
\qquad t\in(-1,0),
\]
and hence $\mathcal{P}(t)$ is strictly decreasing on $(-1,0)$ under the same sufficient condition.

\bibliographystyle{abbrvnat}
\bibliography{References}

@article{kairouz2014extremal,
  title={Extremal mechanisms for local differential privacy},
  author={Kairouz, Peter and Oh, Sewoong and Viswanath, Pramod},
  journal={Advances in neural information processing systems},
  volume={27},
  year={2014}
}

@article{cruyff2008accounting,
  title={Accounting for self-protective responses in randomized response data from a social security survey using the zero-inflated Poisson model},
  author={Cruyff, Maarten JLF and B{\"o}ckenholt, Ulf and van den Hout, Ardo and van der Heijden, Peter GM},
  journal={Annals of Applied Statistics},
  volume={10},
  year={2008}
}

@article{warner1965randomized,
  title={Randomized response: A survey technique for eliminating evasive answer bias},
  author={Warner, Stanley L},
  journal={Journal of the American statistical association},
  volume={60},
  number={309},
  pages={63--69},
  year={1965},
  publisher={Taylor \& Francis}
}

@article{greenberg1969unrelated,
  title={The unrelated question randomized response model: Theoretical framework},
  author={Greenberg, Bernard G and Abul-Ela, Abdel-Latif A and Simmons, Walt R and Horvitz, Daniel G},
  journal={Journal of the American Statistical Association},
  volume={64},
  number={326},
  pages={520--539},
  year={1969},
  publisher={Taylor \& Francis}
}

@article{kuk1990asking,
  title={Asking sensitive questions indirectly},
  author={Kuk, Anthony YC},
  journal={Biometrika},
  volume={77},
  number={2},
  pages={436--438},
  year={1990},
  publisher={Oxford University Press}
}

@book{chaudhuri2020randomized,
  title={Randomized response: Theory and techniques},
  author={Chaudhuri, Arijit and Mukerjee, Rahul},
  year={2020},
  publisher={Routledge}
}

@article{boruch1971assuring,
  title={Assuring confidentiality of responses in social research: a note on strategies},
  author={Boruch, Robert F},
  journal={The American Sociologist},
  pages={308--311},
  year={1971},
  publisher={JSTOR}
}

@article{fox1986randomized,
  title={Randomized response: A method for sensitive surveys},
  author={Fox, James Alan and Tracy, Paul E},
  journal={(No Title)},
  year={1986}
}

@article{blair2015design,
  title={Design and analysis of the randomized response technique},
  author={Blair, Graeme and Imai, Kosuke and Zhou, Yang-Yang},
  journal={Journal of the American Statistical Association},
  volume={110},
  number={511},
  pages={1304--1319},
  year={2015},
  publisher={Taylor \& Francis}
}

@article{dwork2014algorithmic,
  title={The algorithmic foundations of differential privacy},
  author={Dwork, Cynthia and Roth, Aaron and others},
  journal={Foundations and Trends{\textregistered} in Theoretical Computer Science},
  volume={9},
  number={3--4},
  pages={211--407},
  year={2014},
  publisher={Now Publishers, Inc.}
}

@book{chang2023privacy,
  title={Privacy-Preserving Machine Learning},
  author={Chang, J Morris and Zhuang, Di and Samaraweera, G and Samaraweera, G Dumindu},
  year={2023},
  publisher={Simon and Schuster}
}

@inproceedings{dwork2006differential,
  title={Differential privacy},
  author={Dwork, Cynthia},
  booktitle={International colloquium on automata, languages, and programming},
  pages={1--12},
  year={2006},
  organization={Springer}
}

@inproceedings{erlingsson2014rappor,
  title={Rappor: Randomized aggregatable privacy-preserving ordinal response},
  author={Erlingsson, {\'U}lfar and Pihur, Vasyl and Korolova, Aleksandra},
  booktitle={Proceedings of the 2014 ACM SIGSAC conference on computer and communications security},
  pages={1054--1067},
  year={2014}
}

@article{mangat1990alternative,
  title={An alternative randomized response procedure},
  author={Mangat, NS and Singh, Ravindra},
  journal={Biometrika},
  volume={77},
  number={2},
  pages={439--442},
  year={1990},
  publisher={Oxford University Press}
}

@misc{team2017learning,
  title={Learning with Privacy at Scale},
  author={Differential Privacy Team, Apple},
    url = {https://machinelearning.apple.com/research/learning-with-privacy-at-scale},
  year={2017}
}

@article{ding2017collecting,
  title={Collecting telemetry data privately},
  author={Ding, Bolin and Kulkarni, Janardhan and Yekhanin, Sergey},
  journal={Advances in Neural Information Processing Systems},
  volume={30},
  year={2017}
}

@inproceedings{dajani2017modernization,
  title={The modernization of statistical disclosure limitation at the US Census Bureau},
  author={Dajani, Aref N and Lauger, Amy D and Singer, Phyllis E and Kifer, Daniel and Reiter, Jerome P and Machanavajjhala, Ashwin and Garfinkel, Simson L and Dahl, Scot A and Graham, Matthew and Karwa, Vishesh and others},
  booktitle={September 2017 meeting of the Census Scientific Advisory Committee},
  year={2017}
}

@inproceedings{simmons1967unrelated,
  title={The unrelated question randomized response model proceedings in the social statistics section},
  author={Simmons, WR and Horvitz, DG and Shah, BV and others},
  booktitle={American Statistical Association},
  volume={64},
  pages={520--539},
  year={1967}
}

@article{abernathy1970estimates,
  title={Estimates of induced abortion in urban North Carolina},
  author={Abernathy, James R and Greenberg, Bernard G and Horvitz, Daniel G},
  journal={Demography},
  volume={7},
  pages={19--29},
  year={1970},
  publisher={Springer}
}

@article{gingerich2010understanding,
  title={Understanding off-the-books politics: Conducting inference on the determinants of sensitive behavior with randomized response surveys},
  author={Gingerich, Daniel W},
  journal={Political Analysis},
  volume={18},
  number={3},
  pages={349--380},
  year={2010},
  publisher={Cambridge University Press}
}

@article{tezcan1981prevalence,
  title={Prevalence and reporting of induced abortion in Turkey: two survey techniques},
  author={Tezcan, Sabahat and Omran, Abdel R},
  journal={Studies in family planning},
  pages={262--271},
  year={1981},
  publisher={JSTOR}
}

@article{lara2006measure,
  title={The measure of induced abortion levels in Mexico using random response technique},
  author={Lara, Diana and Garc{\'\i}a, Sandra G and Ellertson, Charlotte and Camlin, Carol and Su{\'a}rez, Javier},
  journal={Sociological Methods \& Research},
  volume={35},
  number={2},
  pages={279--301},
  year={2006},
  publisher={Sage Publications Sage CA: Thousand Oaks, CA}
}

@article{lara2004measuring,
  title={Measuring induced abortion in Mexico: a comparison of four methodologies},
  author={Lara, Diana and Strickler, Jennifer and Olavarrieta, Claudia Diaz and Ellertson, Charlotte},
  journal={Sociological Methods \& Research},
  volume={32},
  number={4},
  pages={529--558},
  year={2004},
  publisher={Sage Publications}
}

@article{krumpal2012estimating,
  title={Estimating the prevalence of xenophobia and anti-Semitism in Germany: A comparison of randomized response and direct questioning},
  author={Krumpal, Ivar},
  journal={Social science research},
  volume={41},
  number={6},
  pages={1387--1403},
  year={2012},
  publisher={Elsevier}
}

@article{rosenfeld2016empirical,
  title={An empirical validation study of popular survey methodologies for sensitive questions},
  author={Rosenfeld, Bryn and Imai, Kosuke and Shapiro, Jacob N},
  journal={American journal of political science},
  volume={60},
  number={3},
  pages={783--802},
  year={2016},
  publisher={Wiley Online Library}
}

@article{st2012identifying,
  title={Identifying indicators of illegal behaviour: carnivore killing in human-managed landscapes},
  author={St John, Freya AV and Keane, Aidan M and Edwards-Jones, Gareth and Jones, Lauren and Yarnell, Richard W and Jones, Julia PG},
  journal={Proceedings of the Royal Society B: Biological Sciences},
  volume={279},
  number={1729},
  pages={804--812},
  year={2012},
  publisher={The Royal Society}
}

@article{stubbe2014prevalence,
  title={Prevalence of use of performance enhancing drugs by fitness centre members},
  author={Stubbe, Janine H and Chorus, Astrid MJ and Frank, Laurence E and de Hon, Olivier and van der Heijden, Peter GM},
  journal={Drug testing and analysis},
  volume={6},
  number={5},
  pages={434--438},
  year={2014},
  publisher={Wiley Online Library}
}

@article{warner1986omitted,
  title={The omitted digit randomized response model for telephone applications},
  author={Warner, Stanley L},
  journal={Proceedings of the Social Survey Research Methods Section of the American Statistical Association},
  pages={441--443},
  year={1986}
}

@article{edgell1982validity,
  title={Validity of forced responses in a randomized response model},
  author={Edgell, Stephen E and Himmelfarb, Samuel and Duchan, Karen L},
  journal={Sociological Methods \& Research},
  volume={11},
  number={1},
  pages={89--100},
  year={1982},
  publisher={Sage Publications}
}

@inproceedings{vanderheijden1996some,
  title     = {Some logistic regression models for randomized response data},
  author    = {van der Heijden, Peter G. M. and van Gils, Ger},
  booktitle = {Proceedings of the 11th International Workshop on Statistical Modelling},
  address   = {Orvieto, Italy},
  year      = {1996},
  month     = jul
}

@inproceedings{wang2016using,
  title={Using randomized response for differential privacy preserving data collection.},
  author={Wang, Yue and Wu, Xintao and Hu, Donghui},
  booktitle={EDBT/ICDT Workshops},
  volume={1558},
  pages={0090--6778},
  year={2016}
}

@article{hoglinger2018more,
  title={More is not always better: An experimental individual-level validation of the randomized response technique and the crosswise model},
  author={H{\"o}glinger, Marc and Jann, Ben},
  journal={PloS one},
  volume={13},
  number={8},
  pages={e0201770},
  year={2018},
  publisher={Public Library of Science San Francisco, CA USA}
}

\end{document}